\documentclass[10pt]{article}
\usepackage{amssymb}
\newcommand{\ket}[1]{\left | #1 \right \rangle}

\def\openone{\leavevmode\hbox{\small1\kern-3.8pt\normalsize1}}

\def\cn{{\cal N}}
\def\co{{\cal O}}

\def\ci{{\cal I}}

\def\cc{{\cal C}}
\def\cp{{\cal P}}

\def\mod{\,\, {\rm mod}\,\,}
\def\gcd{{\rm gcd}}
\def\numd{\QQ[\sqrt{d}]}
\def\redid{\ZZ +\frac{b+\sqrt{D}}{2a}\ZZ}

\def\RR{\mathbb{R}}
\def\ZZ{\mathbb{Z}}
\def\QQ{\mathbb{Q}}
\def\NN{\mathbb{N}}

\newtheorem{theorem}{Theorem}
\newtheorem{definition}{Definition}
\newtheorem{lemma}{Lemma}

\newtheorem{proposition}{Proposition}
\newtheorem{corollary}{Corollary}

\newcommand{\mmatrix}[4]{\left( \begin{array}{cc} #1 & #2 \\ #3 &
#4 \end{array} \right)}
\newcommand{\vvec}[2]{\left( \begin{array}{cc} #1  \\ #2 \end{array} \right)}

\newcommand{\beq}{\begin{equation}}
\newcommand{\eeq}{\end{equation}}
\newcommand{\beqa}{\begin{eqnarray}}
\newcommand{\eeqa}{\end{eqnarray}}

\newcommand{\qed}{\mbox{\rule[0pt]{1.5ex}{1.5ex}}}
\newcommand{\poly}{{\rm poly}}

\begin{document}
\begin{center}
{\LARGE\bf Notes on Hallgren's efficient quantum algorithm\\
for solving Pell's equation }\\
\bigskip
{\normalsize Richard Jozsa}\\
\bigskip
{\small\it Department of Computer Science, University of
Bristol,\\ Merchant Venturers Building, Bristol BS8 1UB U.K.} \\[4mm]
\date{today}
\end{center}

\begin{abstract} Pell's equation is $x^2-dy^2=1$ where $d$ is a
square-free integer and we seek positive integer solutions
$x,y>0$. Let $(x_0,y_0)$ be the smallest solution (i.e. having
smallest $A=x_0+y_0\sqrt{d}$). Lagrange showed that every solution
can easily be constructed from $A$ so given $d$ it suffices to
compute $A$. It is known that $A$ can be exponentially large in
$d$ so just to write down $A$ we need exponential time in the
input size $\log d$. Hence we introduce the regulator $R=\ln A$
and ask for the value of $R$ to n decimal places. The best known
classical algorithm has sub-exponential running time $O(\exp
\sqrt{\log d}, \poly(n))$. Hallgren's quantum algorithm gives the
result in polynomial time $O(\poly(\log d),\poly(n))$ with
probability $1/\poly(\log d)$. The idea of the algorithm falls
into two parts: using the formalism of algebraic number theory we
convert the problem of solving Pell's equation into the problem of
determining $R$ as the period of a function on the real numbers.
Then we generalise the quantum Fourier transform period finding
algorithm to work in this situation of an irrational period on the
(not finitely generated) abelian group of real numbers.

These notes are intended to be accessible to a reader having no
prior acquaintance with algebraic number theory; we give a self
contained account of all the necessary concepts and we give
elementary proofs of all the results needed. Then we go on to
describe Hallgren's generalisation of the quantum period finding
algorithm, which provides the efficient computational solution of
Pell's equation in the above sense.
\end{abstract}
\bigskip
\begin{center} {\large\bf CONTENTS}\\[3mm]

\begin{tabular}{rll}
{\bf 1} & The computational task of solving Pell's equation &
\pageref{sect1}
\\ & \hspace{3mm} {\bf 1.1} Approach for the efficient quantum algorithm
& \pageref{effalg} \\ & \hspace{3mm} {\bf 1.2} Note to the reader
&
\pageref{noter} \\
{\bf 2} &  Algebraic integers in a quadratic number field &
\pageref{algint} \\ {\bf 3} & Algebraic integers and Pell's
equation & \pageref{pell-int} \\ {\bf 4} & Ideals of the algebraic
integers & \pageref{ideals} \\  & \hspace{3mm} {\bf 4.1} Principal
ideals and periodicity & \pageref{pridper} \\ {\bf 5} &
Presentations of ideals & \pageref{pres}
\\ {\bf 6} & Reduced ideals and the reduction operator $\rho$ &
\pageref{redsect} \\ & \hspace{3mm} {\bf 6.1} Reduced ideals  & \pageref{redsect1} \\
& \hspace{3mm} {\bf 6.2} The reduction operator $\rho$ & \pageref{redsect2} \\
& \hspace{3mm} {\bf 6.3} The principal cycle of reduced ideals & \pageref{redsect3} \\
& \hspace{3mm} {\bf 6.4} The inverse of the reduction operator & \pageref{redsect4} \\
{\bf 7} & The distance function for ideals & \pageref{distance} \\
& \hspace{3mm} {\bf 7.1} Products
of ideals -- making large distance jumps & \pageref{distance1} \\
{\bf 8} & Summary -- the picture so far & \pageref{summary} \\
{\bf 9} & The
periodic function for Hallgren's algorithm & \pageref{hall-per} \\
{\bf 10} & The quantum algorithm for irrational periods on $\RR$ &
\pageref{quantum} \\ {\bf 11} & Further remarks & \pageref{further} \\
& Acknowledgements & \pageref{ackn}\\ & References &
\pageref{refs}
\end{tabular}

\end{center}
\newpage

\section{The computational task of solving Pell's equation}\label{sect1}
Let $\NN$, $\ZZ$, $\QQ$ and $\RR$ denote respectively the sets of
natural numbers, integers, rational numbers and real numbers.
Pell's equation is \beq \label{pell} x^2 -dy^2=1 \eeq where $d\in
\NN$ is a square-free natural number (i.e. not divisible by $p^2$
for any prime $p$). We wish to find (positive) integer solutions
$(x,y)\in \ZZ\times \ZZ$.

Pell's equation has a rich and colourful history spanning almost
2000 years (cf. \cite{len02,will02} and further references
therein). Diophantus (c. 250AD) gave solutions for $d=26$ and 30
and we read that the English mathematician John Pell (1610-1685)
actually had nothing to do with the equation! The appellation
``Pell's equation'' is based on a confusion originating with Euler
who mis-attributed a solution to Pell which was actually provided
by Lord Brouckner in response to a challenge of Fermat.

It is known that Pell's equation has infinitely many solutions for
any square-free $d$. An elementary proof of this fact can be found
in \cite{ire-ros} chapter 17 \S 5.

Since $\sqrt{d}$ is irrational for any square-free integer $d$, we
have \beq \label{equal} \mbox{ $a+b\sqrt{d}=x+y\sqrt{d}$
\hspace{3mm}for \hspace{3mm} $a,b,x,y \in \ZZ$ (or $\in \QQ$)
\hspace{5mm} {\bf iff} \hspace{5mm} $a=x$ and $b=y$.} \eeq Hence
we can uniquely code any solution $(x,y)$ as $x+y\sqrt{d}\in \RR$.
Correspondingly we will say that $\sigma \in \RR$ is a solution of
Pell's equation if $\sigma$ has the form $\sigma = s+t\sqrt{d}$
with $s,t \in \ZZ$ and $s^2-dt^2=1$.

If $\xi = x+y\sqrt{d}$ for $x,y \in\QQ$ we introduce the
conjugation operation
\[ \overline{\xi} = \overline{x+y\sqrt{d}}= x-y\sqrt{d} \]
which is well defined by eq. (\ref{equal}). It has the following
immediate properties: \beq \label{conjop}
\overline{\overline{\xi}}=\xi \hspace{5mm} \overline{\xi+\eta} =
\overline{\xi}+\overline{\eta} \hspace{5mm} \overline{\xi
\eta}=\overline{\xi}\, \overline{\eta}. \eeq Pell's equation can
be written \beq \label{pell1} \xi \overline{\xi}=1 \eeq so any
solution $\xi=x+y\sqrt{d}$ has the property that
$\overline{\xi}=x-y\sqrt{d}=\frac{1}{\xi}$.

\begin{proposition}\label{num1} If $\alpha=a+b\sqrt{d}$ and
$\xi=x+y\sqrt{d}$ are both solutions of Pell's equation then so
are (i) $\overline{\alpha}$ and (ii) $\alpha \xi= (a+b\sqrt{d})
(x+y\sqrt{d})$ (where we multiply out the RHS and write it as
$s+t\sqrt{d}= (ax+byd)+(ay+bx)\sqrt{d}$).\\ In particular
$\alpha^n = (a+b\sqrt{d})^n$ is a solution for all $n\in \ZZ$.
\end{proposition} \noindent {\bf Proof}\,\,  (i) follows
immediately from eq. (\ref{pell1}) and the definition of
conjugation. For (ii) we note that
$\alpha\overline{\alpha}=\xi\overline{\xi}=1$ so
$(\alpha\xi)(\overline{\alpha\xi})=\alpha\overline{\alpha}\,
\xi\overline{\xi}=1$ i.e. $\alpha\xi$ is a solution. Similarly
$\overline{\alpha^n}=\overline{\alpha}^n$ so $\alpha^n
\overline{\alpha^n} = (\alpha\overline{\alpha})^n =1$. $\qed$

\noindent {\bf Example}\,\, For $d=5$ we have the solution
$9+4\sqrt{5}$ (i.e. $x=9$ and $y=4$). Now we easily check that
$(9+4\sqrt{5})^3 = 2889+1292\sqrt{5}$ so $x=2889$ and $y=1292$ is
also a solution. In fact {\em every} solution for $d=5$ can be
generated as a power of $9+4\sqrt{5}$ (cf. theorem \ref{num2}
below).

\begin{theorem}\label{num2} {\rm (Lagrange 1768)} Let
$\xi_1=x_1+y_1\sqrt{d}$ be the least positive solution of Pell's
equation i.e. $x_1, y_1>0$ and $\xi_1$ is minimum amongst such
solutions. Then every positive solution $(s,t)$ is obtained as a
power of $\xi_1$:
\[ s+t\sqrt{d} = (x_1+y_1\sqrt{d})^n \hspace{5mm} \mbox{for some
$n\in \NN$.} \]
\end{theorem}

\noindent {\bf Proof}\,\,  From proposition \ref{num1} we have
that $\xi_1^n$ is a positive solution for any $n\in \NN$.\\
Conversely suppose that $s+t\sqrt{d}$ is a positive solution not
of the form $\xi^n_1$ for any $n$. Then there exists $k\in \NN$
with \[ \xi_1^k <s+t\sqrt{d}< \xi_1^{k+1} \] so \beq \label{star}
1<(s+t\sqrt{d})(x_1-y_1\sqrt{d})^k < (x_1+y_1\sqrt{d}). \eeq Now
write \[ \alpha= a+b\sqrt{d}=(s+t\sqrt{d})(x_1-y_1\sqrt{d})^k. \]
By proposition \ref{num1} we have $\alpha\overline{\alpha}=1$ so
$(a,b)$ is a solution of Pell's equation and eq. (\ref{star}) says
\[ 1<a+b\sqrt{d}<x_1+y_1\sqrt{d}. \]
To complete the proof we show that $a>0$ and $b>0$ contradicting
the fact that $\xi$ was the {\em least} positive solution. Since
$a-b\sqrt{d}=\frac{1}{a+b\sqrt{d}}$ we have
\[ 0<a-b\sqrt{d}<1. \] From $a+b\sqrt{d}>1$ and $a-b\sqrt{d}>0$ we
get $a>\frac{1}{2}$. Hence $a\geq 1$.\\ From $a-b\sqrt{d}<1$ we
get $b>\frac{a-1}{\sqrt{d}} \geq 0$. Hence both $a$ and $b$ are
$>0$ and we have our contradiction. Thus every positive solution
must be a power of $\xi_1$. $\qed$

The smallest positive solution $\xi(d)=x_1+y_1\sqrt{d}$ is called
the {\bf fundamental solution}. Solving Pell's equation is
equivalent to giving the fundamental solution. As a computational
task the input is the number $d$ with input size $\log d$. However
the magnitude of $\xi(d)$ can be as large as $\xi(d)\sim
O(e^{\sqrt{d}})$ so even to write down $\xi(d)$ we potentially
need $O(\sqrt{d})$ digits which is exponentially large in the
input size.

Examples of fundamental solutions are given in the following
table.
\[ \begin{array}{ccc} d & x_1 & y_1 \nonumber \\
2&3&1 \nonumber \\
3 & 2 & 1 \nonumber \\
5 & 9 & 4 \nonumber \\
13 & 649 & 180 \nonumber \\
14 & 15 & 4 \nonumber \\
15 & 4 & 1 \nonumber \\
29 & 9801 & 1820 \nonumber \\
61 & 1766319049 & 226153980 \nonumber \\
109 & 158070671986249 & 15140424455100 \nonumber \\
2009 & 141012534067201 & 3146065416960 \nonumber \\
4009 & 3799 & 60 \nonumber \\
6009 & \mbox{\tiny 131634010632725315892594469510599473884013975 }
&
\mbox{\tiny 1698114661157803451688949237883146576681644 } \nonumber \\
6013 & 40929908599 & 527831340 \nonumber \\
10209 & 130969496245430263159443178775 &
1296219513663218157975941956 \nonumber \\
16383 & 128 & 1 \nonumber \\
\end{array}
\]
(The interested reader can find more examples at
\[ \mbox{http://www.bioinfo.rpi.edu/$\sim$zukerm/cgi-bin/dq.html} \] and note also
the remark after theorem \ref{num10} below, concerning the
equation $x^2-dy^2=-1$). To get around the exponentially large
size of the integers $x_1$ and $y_1$ we introduce the {\bf
regulator} (an irrational number):
\[ R=\ln (x_1+y_1\sqrt{d}). \]
Then $\lceil R \rceil$ (the least integer $\geq R$) has $O(\log
d)$ digits. \\ Our fundamental computational task becomes:\\ Given
$d$ (a square-free integer) find the regulator $R$ to $n$ digits
of accuracy in poly$(\log d, n)$ time.

The best known classical algorithm has running time
$O(e^{\sqrt{\log d}}{\rm poly}(n))$, which is exponential in the
input size.\\ Actually we can avoid having to consider the
accuracy parameter $n$ using the following result, stating that it
suffices to compute the integer part of $R$.
\begin{proposition} \label{num3} If we are given the closest
integer above or below the regulator $R$ then there exists a
classical algorithm that will compute $R$ to $n$ digits of
accuracy with running time poly$(n,\log d)$. \end{proposition} We
will give a proof later (in section \ref{hall-per} after the proof
of theorem \ref{num36}).
\subsection{Approach for the efficient quantum
algorithm} \label{effalg} Using results from algebraic number
theory we will set up a function $h:\RR \rightarrow A$ (where the
nature of the set $A$ will be given later, but for now,
think of $A$ as being $\RR$ too) with the following properties. \\
(a) $h$ is computable in polynomial time. More precisely if $x$ is
a real number which is an integer multiple of $10^{-n}$ then the
value of $h(x)$ can
be computed accurate to $10^{-n}$ in poly$(\log d, \log x,n)$ time. \\
(b) $h$ is
periodic on $\RR$ with (irrational) period $R$, the regulator,
and $h$ is one-to-one within each period.\\
We then adapt the standard quantum Fourier transform  period
finding algorithm (that is used in Shor's algorithm and other
hidden subgroup problems) to work in the case of the (not finitely
generated) abelian group $\RR$ and irrational period $R$. In fact
we will just discretise $h$ by taking $x=k/N$ for suitably large
chosen values of $N$, and round off values of $h(x)$ too, to get a
discrete domain and range. Then we show that the resulting
function (which is not quite periodic because of rounding effects
in both the domain and range) can give the desired approximations
to $R$ to increasing accuracy (as $N$ is increased).

Thus the ingredients of Hallgren's algorithm fall into two
essentially disjoint parts. The first part constructs the function
$h$ from the classical mathematics of algebraic number theory and
shows that it is efficiently computable. This part has no quantum
ingredients. The second part (having the whole quantum content)
shows how to generalise the standard quantum period finding
algorithm to work on real numbers, to determine an irrational
period to any desired accuracy.

\subsection{Note to the reader} \label{noter}
One of the most interesting features of Hallgren's algorithm is
that it expands the applicability of quantum computation into new
areas of mathematics {\em viz.} fundamental computational problems
of algebraic number theory, especially the study of ideals of the
algebraic integers in quadratic number fields (cf. later for an
explanation of all these terms). In particular we get efficient
quantum algorithms for the solution for Pell's equation, the
principal ideal problem and the determination of the class group
(and none of these have known classical efficient solutions).

In these notes we assume no prior knowledge of algebraic number
theory. After much lucubration we have developed a self contained
account of the necessary parts of this theory with all properties
and theorems being proved by elementary means.  Nevertheless for
readers unacquainted with algebraic number theory it may be
advisable to skip many of the proofs on initial reading, while
focussing on the statements, concepts and terminology.

Our account of algebraic number theory is based primarily on
\cite{btw} with further reference to \cite{coh93}, \cite{len82},
\cite{ire-ros} and \cite{will02}. The description of the
generalised quantum period finding algorithm and its application
to Pell's equation is just an expanded version of Hallgren's
account in \cite{hallgren}.

\section{Algebraic integers in a quadratic number field}
\label{algint} Let $d$ be a square-free positive integer. The {\bf
quadratic number field} $\QQ[\sqrt{d}]$ is defined to be the set
\[ \QQ[\sqrt{d}] = \{ r_1+r_2\sqrt{d}: r_1,r_2 \in \QQ \}. \]
We think of $\QQ[\sqrt{d}]$ as an extension of the usual rational
numbers $\QQ$. (Indeed the irrational $\sqrt{d}$ amongst rationals
behaves rather like the complex $i$ amongst reals). It is clearly
closed under the usual arithmetic operations of addition,
multiplication and formation of reciprocals e.g. if
$\xi=r_1+r_2\sqrt{d}\neq 0$ then $\frac{1}{\xi}=
\overline{\xi}/(\xi \overline{\xi}) =
(r_1-r_2\sqrt{d})/(r_1^2-r_2^2d)$ which is clearly in
$\QQ[\sqrt{d}]$ (and the denominator is never zero as $\sqrt{d}$
is irrational). Also in addition to eq. (\ref{conjop}) we have
$\overline{\xi/\eta}=\overline{\xi}/\overline{\eta}$.

$\xi \in \QQ[\sqrt{d}]$ is an {\bf algebraic integer} if $\xi$ is
the root of a polynomial with {\em integer} coefficients {\em and
with leading coefficient} 1 i.e. for some $n\in \NN$ we have
$\xi^n +a_{n-1}\xi^{n-1} + \ldots +a_1\xi+a_0=0$ where $a_0,a_1,
\ldots ,a_{n-1} \in \ZZ$. Let $\co$ denote the set of all
algebraic integers in $\QQ[\sqrt{d}]$. $\co \subset \QQ[\sqrt{d}]$
is a generalisation of the usual notion of integers $\ZZ
\subset\QQ$ in the following sense.

\begin{proposition} \label{num4} In $\QQ$ the algebraic integers are
just the usual integers $\ZZ$. \end{proposition}

\noindent {\bf Proof}\,\,  Suppose that $\frac{p}{q}$ (with $p,q
\in \NN$ having no common divisors except 1) satisfies a
polynomial equation as above:
\[ (\frac{p}{q})^n + a_{n-1}(\frac{p}{q})^{n-1} + \ldots +a_0 =0.\]
Then multiplying through by $q^{n-1}$ shows that $\frac{p^n}{q}$
is an integer i.e. $q$ must divide $p$ so $q=\pm 1$. $\qed$

\begin{proposition} \label{num4.5} If $\alpha_1$ and $\alpha_2$
are algebraic integers in $\QQ[\sqrt{d}]$ then $\alpha_1+\alpha_2$
and $\alpha_1\alpha_2$ are also algebraic
integers.\end{proposition}

\noindent {\bf Proof}\,\,  Clearly $\alpha_1+\alpha_2$ and
$\alpha_1\alpha_2$ are in $\QQ[\sqrt{d}]$. Since $\alpha_1$ and
$\alpha_2$ are algebraic integers there are polynomial equations
\[ \begin{array}{l}
\alpha_1^{n} + k_{n-1}\alpha_1^{n-1} + \ldots +k_0 =0 \\
\alpha_2^{m} + l_{m-1}\alpha_2^{m-1} + \ldots +l_0 =0 \\
\mbox{with $k_i, l_i \in \ZZ$}. \end{array} \] Let $V$ be the set
of all $\ZZ$-linear combinations of
$\beta_{ij}=\alpha_1^i\alpha_2^j$ for $0\leq i<n$ and $0\leq j<m$
i.e. $V=\{ \sum k_{ij}\beta_{ij}: k_{ij}\in \ZZ \}$. Clearly
$\gamma\in V$ implies that $\alpha_1\gamma \in V$ and $\alpha_2
\gamma \in V$ (as we can use the above polynomial equations to
express $\alpha_1^n$ and $\alpha_2^m$ in terms of lower powers).
Hence $(\alpha_1+\alpha_2)\gamma$ and $(\alpha_1\alpha_2)\gamma$
are both in $V$. Thus by lemma \ref{num4.75} below, $\alpha_1+
\alpha_2$ and $\alpha_1\alpha_2$ are algebraic integers. $\qed$

\begin{lemma} \label{num4.75} Let $\gamma_1, \ldots ,\gamma_l$ be
any chosen complex numbers and let $V= \{ \sum k_i \gamma_i: k_i
\in \ZZ \}$. Suppose that a complex number $\alpha$ has the
property that $\alpha\gamma \in V$ for all $\gamma \in V$. Then
$\alpha$ is an algebraic integer. \end{lemma}

\noindent {\bf Proof}\,\,  $\alpha\gamma_i \in V$ so
$\alpha\gamma_i = \sum a_{ij}\gamma_j$ for some $a_{ij}\in \ZZ$.
Hence $0=\sum (a_{ij}-\delta_{ij}\alpha) \gamma_j$ so by standard
linear algebra we have ${\rm det}(a_{ij}-\delta_{ij}\alpha)=0$,
giving the required polynomial equation with $\alpha$ as a root.
$\qed$

\begin{proposition}\label{num5} $\xi=r+s\sqrt{d}\in \QQ[\sqrt{d}]$
is an algebraic integer {\bf iff} $2r$ and $r^2-s^2d$ are both
integers.
\end{proposition}

\noindent {\bf Proof}\,\,  Note that $2r=\xi+\overline{\xi}$ and
$r^2-s^2d=\xi\overline{\xi}$. Hence if both of these are integers
then the polynomial equation $(x-\xi)(x-\overline{\xi})=0$  shows
that the root $\xi$ is an algebraic integer.\\ Conversely suppose
that $\xi$ is an algebraic integer. Then $\overline{\xi}\in \co$
too (satisfying the same polynomial equation). Hence by
proposition \ref{num4.5} we have that $2r=\xi+\overline{\xi}$ and
$r^2-s^2d=\xi\overline{\xi}$ are algebraic integers. But these are
both pure rationals  so by proposition \ref{num4} they must be
integers. $\qed$

\noindent {\bf Example}\,\, If $m,n\in \ZZ$  then $m+n\sqrt{d}$ is
always an algebraic integer for any $d$ (as follows immediately
from proposition \ref{num5}) but for some $d$ values there are
more algebraic integers e.g. if $d=5$ then $\frac{1+\sqrt{5}}{2}$
is an algebraic integer (since it is the root of $x^2-x+1=0$).
Below we will explicitly characterise $\co$ for any $d$.

\noindent {\bf Remark on notations} For any $\alpha,\beta
\in\QQ[\sqrt{d}]$ we will write\\ $\ZZ[\alpha] = \{
m+n\alpha:m,n\in \ZZ \}$ \\ $\alpha \ZZ = \{ n\alpha : n\in \ZZ
\}$
\\ $\alpha \ZZ +\beta\ZZ = \{ n\alpha +m\beta: m,n\in \ZZ
\}$ \\ In particular we can write $\ZZ[\alpha]= \ZZ +
\alpha\ZZ$.\\
We will also use the following properties:\\ (a) $\alpha\ZZ =
-\alpha\ZZ$. \\ (b) If $a,b\in \ZZ$ then $a\ZZ+\frac{b}{2}\ZZ =
a\ZZ +\frac{b'}{2}\ZZ$ for any $b'$ having $b'\equiv b \mod 2a$.\\
(c) If $a,b\in \NN$ then Euclid's algorithm states that the
greatest common divisor $\gcd(a,b)$ of $a$ and $b$ can be
expressed as $\gcd(a,b)=ka+lb$ for some $k,l\in\ZZ$. Hence
$a\ZZ+b\ZZ=\gcd(a,b)\ZZ$. (To see this let $g=\gcd(a,b)$. Then LHS
$\subseteq$ RHS as $a=a'g$ and $b=b'g$ for $a',b'\in \ZZ$, so
anything in LHS is a multiple of $g$. Conversely RHS $\subseteq$
LHS, as by Euclid's algorithm, {\em every} multiple of $g$ has the
form $k'a+l'b$ with $k',l'\in \ZZ$).

\begin{theorem}\label{num7} In $\numd$ the set of algebraic integers
has the form \[ \co=\{ m+n\omega :m,n\in \ZZ \} \] where \\ if
$d\equiv 1 \mod 4$ we can take $\omega = \frac{-1+\sqrt{d}}{2}$;\\
if $d\equiv \mbox{2 or 3} \mod 4$ we can take $\omega=\sqrt{d}$.
\\ (Since $d$ is square free we never have $d\equiv 0 \mod 4$).
\end{theorem}

\noindent {\bf Proof}\,\,  Let $\xi=r+s\sqrt{d}\in \numd$. Suppose
$\xi \in \co$. Then $2r\in \ZZ$ and $r^2-s^2d\in \ZZ$ so
$4s^2d=(2r)^2-4(r^2-s^2d) \in \ZZ$. Hence any prime $p>2$ in the
denominator of $s$ must have $p^2$ dividing $d$, which is
impossible as $d$ is square-free. Hence $2s\in \ZZ$. Set $2r=m$
and $2s=n$. Then $r^2-s^2d\in \ZZ$ implies $m^2-dn^2 \equiv 0 \mod
4$. Recall that any square is congruent to 0 or 1 $\mod 4$
(because $(2k)^2 =4k^2 \equiv 0$ and $(2k+1)^2 =
4k^2+4k+1 \equiv 1 \mod 4$). \\
If $d\equiv \mbox{2 or 3}\mod 4$: then $m^2-dn^2 \equiv m^2+2n^2$
or $m^2+n^2 \mod 4$. The only way that $m^2+2n^2$ or $m^2+n^2$ can
be divisible by 4 is for both $m$ and $n$ to be even, and this is
the case iff $r$ and $s$ are integers. Hence if $d\equiv \mbox{2
or 3}\mod 4$ then any algebraic integer has the form $m+n\sqrt{d}$
for $m,n\in \ZZ$. Conversely by proposition \ref{num5} any
expression of this form is an algebraic integer.\\
If $d\equiv 1 \mod 4$: then $m^2-dn^2\equiv m^2-n^2 \mod 4$ and
this can be divisible by 4 iff $m$ and $n$ are both even or both
odd. Thus any algebraic integer has the form
$\frac{m+n\sqrt{d}}{2}$ with $m,n\in \ZZ $ both even or both odd.
Conversely any number $\xi$ of this form has
$\xi+\overline{\xi}\in \ZZ$ and $\xi\overline{\xi}=
\frac{m^2-dn^2}{4}\in \ZZ$ since $d\equiv 1 \mod 4$. Thus by
proposition \ref{num5} $\xi$ is an algebraic integer. Finally
writing
\[ \frac{m+n\sqrt{d}}{2}=\frac{m+n}{2}+n(\frac{-1+\sqrt{d}}{2}) \]
with $\frac{m+n}{2}\in \ZZ$ (as $m,n$ have the same parity) we see
that
\[ \{ \frac{m+n\sqrt{d}}{2}:m,n\in \ZZ\mbox{ are both even or
odd}\} = \{ k+l(\frac{-1+\sqrt{d}}{2}): k,l\in \ZZ \} \hspace{5mm}
\qed
\]
Note that 1 and $\omega$ are linearly independent over $\QQ$ i.e.
$r_1 1+r_2\omega=0$ for $r_1,r_2\in \QQ$ iff $r_1=r_2=0$. Thus we
see that $\co$ behaves like a 2 dimensional ``vector space'' where
1 and $\omega$ are basis vectors but the coefficients for linear
combinations must be integers. Such a structure is called a
$\ZZ$-module. Although we are drawing an analogy with vector
spaces it is interesting to note that actually, the notion of
$\ZZ$-module is the same as the notion of abelian group: clearly
any $\ZZ$-module as above is an abelian group under addition.
Conversely if $(G,+)$ is any abelian group then writing
$g+g+\ldots +g$ ($n$ times) as $(n)g$ and the inverse $-g$ of $g$
as $(-1)g$, then $G$ can be viewed naturally as a $\ZZ$-module
i.e. $G$ is the same as the set of all $\ZZ$-linear combinations
of elements of $G$.

A pair of elements $\alpha,\beta \in \co$ is called an {\bf
integral basis} of $\co$ if $\co = \{ m\alpha+n\beta: m,n\in \ZZ
\}$. Hence $\{ 1, \omega \}$ is an integral basis but the choice
is not unique (cf. proposition \ref{num15} later).
\begin{proposition}\label{num8}
If $\{ \alpha, \beta \}$ is any integral basis of $\co$ then
\[ D=\left| \begin{array}{cc} \alpha & \beta \nonumber \\
\overline{\alpha} & \overline{\beta} \nonumber \end{array}
\right|^2
\]
is a positive integer independent of the choice of basis. $D$ is
called the {\bf discriminant} of $\numd$. In fact if $d\equiv 1
\mod 4$ then $D=d$ and if $d\equiv 2,3 \mod 4$ then $D=4d$.
\end{proposition}

\noindent {\bf Proof}\,\,  Let $\{\alpha,\beta \}$ be an integral
basis. Since $\{1,\omega \}$ is an integral basis we have
\[ \left( \begin{array}{cc} \alpha & \beta \end{array} \right)=
\left( \begin{array}{cc} 1 & \omega \end{array} \right) \left(
\begin{array}{cc} a_1 & a_2 \\ b_1 & b_2 \end{array} \right)
\equiv \left( \begin{array}{cc} 1 & \omega \end{array} \right) M
\] for some integers $a_1,a_2,b_1,b_2$. Hence \[ \left(
\begin{array}{cc} 1 & \omega \end{array} \right) = \left(
\begin{array}{cc} \alpha & \beta \end{array} \right) M^{-1} \]
but $\{ \alpha,\beta \}$ is an integral basis too so $M^{-1}$ must
have integer entries. Since $M$ and $M^{-1}$ both have integer
entries it follows that $\det M =\pm 1$ (because $\det M $ and
$\det M^{-1}= 1/\det M$ are both integers). Thus \[  \left|
\begin{array}{cc} \alpha  & \beta \\ \overline{\alpha} &
\overline{\beta} \end{array} \right|^2 =  \left| \left(
\begin{array}{cc} 1&\omega \\ 1&\overline{\omega} \end{array}
\right) M\right|^2 =| M |^2  \left|
\begin{array}{cc} 1&\omega \\ 1&\overline{\omega} \end{array}
\right|^2 =   \left|
\begin{array}{cc} 1&\omega \\ 1&\overline{\omega} \end{array}
\right|^2. \] Using the explicit values of $\omega$ in theorem
\ref{num7} we get:\\ if $d\equiv 1 \mod 4$ then
$D=d$;\\
if $d\equiv 2,3 \mod 4$ then $D=4d$. $\qed$

Hence an integer $D\in\NN$ is the discriminant of some $\numd$ iff
$D\equiv 1 \mod 4$ or else $D\equiv 0 \mod 4$ and then
$\frac{D}{4}\equiv 2,3 \mod 4$. Using the notion of discriminant
we can give a unified description of $\co$ (without having to
separate two cases of $d\mod 4$). Indeed it will often be easier
to work with $D$ rather than $d$.
\begin{proposition}\label{num9} Let $D$ be the discriminant of
$\numd$. Then in all cases
$\co=\ZZ[\frac{D+\sqrt{D}}{2}]$.\end{proposition}

\noindent {\bf Proof}\,\,  This follows immediately from the above
values of $D$ and $\omega$ for the various cases of $d\mod 4$.
$\qed$

\section{Algebraic integers and Pell's equation}\label{pell-int}

We now establish the connection between algebraic integers and
Pell's equation.

An algebraic integer $\xi\in \co$ is called a {\bf unit} if it has
a multiplicative inverse {\em that is also an algebraic integer}.
 According to this definition, in the
usual integers $\ZZ$ (or rationals $\QQ$), there are only two
units {\em viz.} $\pm1$.
\begin{proposition}\label{num9.1} $\xi=x+y\sqrt{d}\in \co$ is a
unit {\bf iff} $2x\in \ZZ$ and $x^2-dy^2=\pm 1$ (i.e. we have a
strengthening of the conditions in proposition \ref{num5})
.\end{proposition}

\noindent {\bf Proof}\,\,  Recall that for any $\xi\in \numd$
\[ \frac{1}{\xi}= \frac{\overline{\xi}}{\xi\overline{\xi}}
= \frac{x-y\sqrt{d}}{x^2-dy^2}.
\] By
proposition \ref{num5} $\xi\in \co$ iff $2x\in \ZZ$ and
$x^2-dy^2\in \ZZ$ and for $\frac{1}{\xi}$ to be in $\co$ we also
require $\frac{2x}{x^2-dy^2}\in \ZZ$ and
$\frac{x^2-dy^2}{(x^2-dy^2)^2}= \frac{1}{x^2-dy^2}\in \ZZ$ i.e.
$x^2-dy^2 =\pm 1$. $\qed$

Hence (in contrast to $\ZZ$) there are infinitely many units in
$\co$ and they are intimately related to solutions of Pell's
equation i.e. solutions of Pell's equation can be viewed as a
natural generalisation of the fundamental integers $\pm 1$.

\begin{proposition}\label{num9.2} If $a+b\sqrt{d}>1$ is a unit
then $a>0$ and $b>0$.\end{proposition}

\noindent {\bf Proof}\,\,  Arguing by contradiction, suppose that
$\xi =x-y\sqrt{d}>1$ is a unit with $x>0,y>0$. Then \[
0<\frac{1}{\xi}=\frac{x+y\sqrt{d}}{x^2-dy^2}<1. \] But $x^2-dy^2
=\pm 1$ so $\frac{1}{\xi}>0$ gives $x^2-dy^2=1$ and then
$x+y\sqrt{d}<1$. Now $x^2-dy^2=1$ implies  $x\geq1$ in
contradiction to $x+y\sqrt{d}<1$. Similarly $-x+y\sqrt{d}$ cannot
be a unit if it is $>1$. $\qed$

A slight generalisation of theorem \ref{num2} characterises all
units:
\begin{theorem}\label{num10} Let $\epsilon_0$ be the smallest
unit in $\co$ that is greater than 1. Then the set of all units is
given by $\{ \pm \epsilon_0^k$: $k\in \ZZ \}$.
\end{theorem}
\noindent $\epsilon_0$ is called the {\bf fundamental unit}.

\noindent {\bf Proof}\,\,  The proof is very similar to that of
theorem \ref{num2} and we omit duplicating the details. $\qed$

\noindent {\bf Remark}\,\, If $d\equiv 2,3 \mod 4$ then $\co =
\ZZ[\sqrt{d}]$ so if the fundamental unit $\epsilon_0=m+n\sqrt{d}$
has $m^2-dn^2=1$ then the units are exactly the solutions of
Pell's equation. If $m^2-dn^2=-1$ then
$\epsilon_1=\epsilon_0^2=m'+n'\sqrt{d}$ with $m'=2m^2+1$ and
$n'=2mn$ has $m'^2-dn'^2=\epsilon_0^2 \overline{\epsilon}_0^2 =
(-1)^2= 1$ and it generates all
solutions of Pell's equation via theorem \ref{num2}.\\
 If $d\equiv 1
\mod 4$ then $\co = \ZZ[\frac{-1+\sqrt{d}}{2}]$ so some units may
have rational coefficients $r_1+r_2\sqrt{d}$ with denominator 2
(e.g. $d=5$ has $\frac{1\pm \sqrt{5}}{2}$ as units). But given the
fundamental unit we can generate all units in numerical order as
powers and select the smallest unit $\epsilon =x+y\sqrt{d}>1$
having $x^2-dy^2=1$ and $x,y\in \ZZ$. Then via theorem \ref{num2}
we can again generate all solutions of Pell's equation. $\qed$

Hence solving Pell's equation is equivalent to finding the
fundamental unit of the algebraic integers in $\numd$. We define
the {\bf regulator} $R$ of $\co$ by
\[ R=\ln \epsilon_0
\] and our task is now to compute an $n$ digit
approximation to $R$ in poly$(\log d,n)$ time. To define our basic
function $h$ with period $R$ we will need the concept of an ideal
of $\co$ in $\numd$ and more specifically, the notion of reduced
principal ideals.

\section{Ideals of the algebraic integers}\label{ideals}
If $A$ and $B$ are subsets of $\co$ or $\numd$ we define the
product $A\cdot B$ to be the additive span of all products $ab$
with $a\in A$ and $b\in B$ \[ A\cdot B = \{ a_1b_1+ \ldots +a_nb_n
: a_1, \ldots ,a_n \in A, b_1, \ldots ,b_n \in B, n\in \NN \}. \]
\begin{definition} \label{num11} $I\subseteq \co$ is an {\bf
(integral) ideal} of $\co$ if $I\cdot\co=I$ and $\alpha, \beta \in
I$
implies $m\alpha+n\beta \in I$ for all $m,n\in\ZZ$.\\
$I\subseteq \numd$ is a {\bf fractional ideal} of $\co$ if
$I\cdot\co=I$ and $\alpha, \beta \in I$ implies $m\alpha+n\beta
\in I$ for all $m,n\in\ZZ$. \end{definition} In other words an
integral (respectively fractional) ideal of $\co$ is a subset of
$\co$ (respectively $\numd$) that is closed under forming
$\ZZ$-linear combinations of its elements and also closed under
multiplication by elements of $\co$. In standard algebra the basic
object of study ($\co$ here) is generally not embedded in an
ambient structure ($\numd$ here) and the term `ideal' corresponds
to `integral ideal' above.

\noindent {\bf Example} In  $\QQ$ we have $\co=\ZZ$ (the usual
integers) and $I_m = m\ZZ = \{ 0,\pm m,\pm 2m,\ldots \}\subseteq
\ZZ$ is an integral ideal for each $m$ as can easily be verified.
Similarly for $\co \subset\numd$ if $\alpha\in \co$ is any chosen
element then $I_\alpha = \alpha\co= \{ \alpha\xi:\xi\in \co \}$ is
always an integral ideal (or fractional ideal if more generally
$\alpha\in \numd$). It can be shown that for $\ZZ$ the $I_m$'s are
the {\em only} integral ideals whereas for $\co\subset \numd$,
generally the $I_\alpha$'s do {\em not} exhaust all possible
ideals.

\subsection{Principal ideals and periodicity} \label{pridper}
\begin{definition} \label{num12} If $\gamma\in \co$ (respectively
$\numd$) then the set $\gamma\co = \{ \gamma\xi:\xi\in\co \}$ is
always an integral (respectively fractional) ideal and ideals of
this form are called {\bf principal} ideals.\end{definition}
\begin{proposition}\label{num13} $\alpha\co=\beta\co$ {\bf iff}
$\alpha=\beta\epsilon$ where $\epsilon$ is a unit in
$\numd$.\end{proposition}

\noindent {\bf Proof}\,\,  If $\epsilon$ is a unit then it is easy
to see that $\epsilon\co=\co$. Hence if $\alpha=\beta\epsilon$
then $\alpha\co=\beta\epsilon\co=\beta\co$. Conversely suppose
that $\alpha\co=\beta\co$. Since $1\in \co$ we have $\alpha\in
\alpha\co=\beta\co$ so there is $\eta_1\in \co$ with
$\alpha=\beta\eta_1$. Interchanging roles of $\alpha$ and $\beta$
we get $\eta_2\in \co$ with $\beta=\eta_2\alpha$ so
$\alpha=\beta\eta_1=\alpha\eta_2\eta_1$ i.e. $\eta_2\eta_1=1$ and
$\eta_1,\eta_2$ are units. $\qed$

Proposition \ref{num13} with theorem \ref{num10} provides the key
to converting our basic task (of computing regulators) into a
periodicity problem. By theorem \ref{num10} we know that
$\epsilon$ is a unit iff $\epsilon=\epsilon_0^k$ where
$\epsilon_0$ is the fundamental unit. Let $\cp\ci =\{
\xi\co:\xi\in \numd\}$ be the set of all fractional principal
ideals. Thus if we consider the ideal $\xi\co \in \cp\ci$ as a
function of $x=\ln \xi$ i.e. \[ g(x)=e^x \co, \] then $g$ will be
a periodic function with our desired period $R=\ln \epsilon_0$.
However the {\em direct} use of this function would appear to be
computationally problematic for various reasons. In order to
compute $g$ and see the periodicity we would need to be able to
determine that $e^x\co$ and $e^{x'}\co$ are the same ideal when
$x'=x+R$ and furthermore we would need to compute $g$ (in quantum
superposition) for values of $x\sim O(R)$ or larger. In that case,
even the integer part of $e^x$ (being $O(\epsilon_0)\sim
O(e^{\sqrt{d}})$) would have exponentially many digits.  Thus the
arithmetic operations needed to see that $e^x\co = e^{x+R}\co$
would presumably require exponential time $O(\poly (\log d,\log
e^x)) = O(\poly (d))$. Furthermore the sets $\QQ[\sqrt{d}]$ and
$\co$ are dense in $\RR$ and we have infinitely many distinct
(integral) ideals. Thus the identification of the ideal
$\alpha\co$ would generally depend on $\alpha$ to full (infinite)
precision (or alternatively we would need to formulate some
effectively computable notion of two ideals $I$ and $I'$ being
``almost the same''.)

To get around these difficulties we will utilise the notion of
{\em reduced} principal ideals $I$ (a concept which already
appears in Gauss' {\em Disquisitiones Arithmeticae} of 1801) and a
notion of distance $\delta(I)$ of $I$ from the unit ideal $\co$
(which was introduced by D. Shanks in 1972 \cite{shanks}). The set
$\cp\ci_{\rm red}$ of reduced principal ideals will be a {\em
finite} set (although exponentially large in $\log d$). Each
reduced ideal will have a $\poly \log d$ sized description so we
avoid the above problems of infinite precision. Furthermore we
will have an (efficiently computable) operation $\rho :
\cp\ci_{\rm red} \rightarrow \cp\ci_{\rm red}$ which allows us to
cycle through the set of reduced ideals in order of increasing
distance from $\co$ and we will have an effective means of jumping
by exponentially large distances. With each successive application
of these processes (starting on $\co$) we will be able to
efficiently compute the distance increment and the accumulated
distance passes through $R$ as the ideal cycle returns to $\co$.
Once these, and some further ingredients are in place we will be
able to define an efficiently computable function on $\RR$ with
period $R$ and apply the quantum period-finding algorithm
(suitably generalised for irrational periods on $\RR$).

\section{Presentations of ideals}\label{pres} \begin{proposition}
\label{num14} Any principal fractional ideal $I$ has the form \[
I=\alpha\ZZ+\beta\ZZ =\{ m_1\alpha+m_2\beta: m_1,m_2\in\ZZ
\}\hspace{1cm}  \] where $\alpha,\beta \in \numd $ are linearly
independent over $\QQ$.
\end{proposition}

\noindent {\bf Proof}\,\,  Since $\co=\ZZ+\omega\ZZ$, for
$\gamma\co$ we can take $\alpha=\gamma$ and $\beta=\gamma\omega$.
Furthermore $1,\omega $ are linearly independent over $\QQ$ so
$\alpha, \beta$ must be too. $\qed$

\noindent {\bf Remark} Proposition \ref{num14} is actually true
for {\em any} (not necessarily principal) fractional ideal of
$\co$ but we will not need this more general fact.

Thus intuitively any ideal is like a ``2-dimensional vector space
over $\ZZ$'' with the extra property of being closed under
multiplication by $\co$ (which restricts the possible choices of
$\alpha$ and $\beta$). Any set $\{\alpha,\beta \}$ in proposition
\ref{num14} is called an {\bf integral basis} of the fractional
ideal. Changes of basis must respect the restriction that
coefficients are required to be integers.
\begin{proposition}\label{num15} Let $\{ \alpha,\beta \}$ be an
integral basis of a fractional ideal $I$. Then $\{
\alpha',\beta'\}$ is another integral
basis {\bf iff} \beq\label{em} \left( \begin{array}{c} \alpha' \\
\beta'
\end{array} \right) = M\left( \begin{array}{c} \alpha \\ \beta
\end{array} \right) \eeq where $M$ is a $2\times 2$ matrix with
 integer entries and $ \det M
=\pm1$.\end{proposition}

\noindent {\bf Proof}\,\,  ($\Rightarrow$) Since
$\alpha',\beta'\in I$, eq. (\ref{em}) must hold for some matrix
$M$ with integer entries. By moving $M$ to LHS as $M^{-1}$ we see
that $M^{-1}$ must also have integer entries so (just as in
proposition \ref{num8}) we must have $\det M=\pm 1$.\\
($\Leftarrow$) Conversely if eq. (\ref{em}) holds with $\det M=\pm
1$ then each of $\alpha\ZZ+\beta\ZZ$ and $\alpha'\ZZ+\beta'\ZZ$ is
contained in the other since the two sets $\{\alpha,\beta \}$ and
$\{ \alpha',\beta'\}$ are related as linear combinations with
integer coefficients. $\qed$

 If $\{ \alpha,\beta \}$ is any
integral basis of a fractional ideal $I$ then we introduce the
absolute value
\[ \cn (I) = \left| \det \left( \begin{array}{cc} \alpha & \beta \\
\overline{\alpha} & \overline{\beta} \end{array} \right) \right|
/\sqrt{D} \]
\begin{proposition}\label{num16}
$\cn(I)$ is independent of the choice of integral basis. If
$I=\gamma \co$ is a principal integral ideal then $\cn(I)$ is the
integer $|\gamma\overline{\gamma}|$. \end{proposition}

\noindent {\bf Proof}\,\,  If $\{ \alpha',\beta' \}$ is any other
integral ideal then $(\alpha'\hspace{2mm} \beta')= (\alpha
\hspace{2mm} \beta ) M$ where $\det M=\pm 1$. Hence \[ \det \left(
\begin{array}{cc} \alpha' & \beta' \\ \overline{\alpha'} &
\overline{\beta'} \end{array} \right) = \det M \det \left(
\begin{array}{cc} \alpha & \beta \\ \overline{\alpha} &
\overline{\beta} \end{array} \right)= \pm \det \left(
\begin{array}{cc} \alpha & \beta \\ \overline{\alpha} &
\overline{\beta} \end{array} \right). \] If $I=\gamma \co$ we can
take $\alpha=\gamma$ and $\beta = \gamma (D+\sqrt{D})/2$ and
compute $\cn(I)$ directly giving $|\gamma\overline{\gamma}|$ which
is an integer for $\gamma \in \co$ by proposition \ref{num5} .
$\qed$
\begin{proposition}\label{num17} Any fractional ideal $I$ has an
integral basis $\{ \alpha,\beta\}$ with $\alpha>0$ rational.
Furthermore $\alpha$ is uniquely determined as the least positive
rational in $I$. If $I$ is an integral ideal then $\alpha$ is an
integer (and so $I$ contains no rationals that are not
integers).\end{proposition}

\noindent {\bf Proof}\,\,  If $\alpha = r_1+r_2\sqrt{d},
\beta=s_1+s_2\sqrt{d}$ with $r_1,r_2,s_1,s_2 \in \QQ$ is any
integral basis there exist integers $m,n$ such that $mr_2+ns_2=0$
and $\gcd(m,n)=1$. Thus by Euclid's algorithm there exist integers
$a,b$ such that $am-bn=\gcd(m,n)=1$ i.e. the matrix
\[ M=\mmatrix{m}{n}{a}{b} \] has $\det M=1$. Then the integral
basis \[ \vvec{\alpha'}{\beta'} = M \vvec{\alpha}{\beta}
\] has $\alpha'$ rational (which we may take to be $>0$ by a
change of sign). Since $\{ \alpha',\beta'\}$ is linearly
independent over $\QQ$, $\beta'$ cannot be rational. Any element
$\xi$ of $I$ can be written $\xi=m\alpha'+n\beta'$ so $\xi$ is
rational iff $n=0$. Thus $\alpha'$ is the least positive rational
in $I$. \\ If $I$ is an integral ideal then $I\subseteq
\co=\ZZ[\frac{D+\sqrt{D}}{2}]= \{ m+n\frac{D+\sqrt{D}}{2}: m,n\in
\ZZ \}$ where $\sqrt{D}$ is irrational. Hence the only rationals
in $I$ are integers so $\alpha'$ must be an integer. $\qed$

\begin{proposition}\label{num18} $I\subseteq \numd$ is a
fractional ideal {\bf iff} there is $m\in \NN$ such that $mI$ is
an integral ideal (hence justifying the terminology `fractional
ideal'). \end{proposition}

\noindent {\bf Proof}\,\,  $(\Leftarrow)$ If $mI$ is an integral
ideal
the clearly $I=\frac{1}{m}(mI)$ is a fractional ideal.\\
$(\Rightarrow)$ Let $\{ \alpha,\beta \}$ be an integral basis of
$I$ and let $k$ be any integer such that $k\alpha,k\beta \in
\ZZ[\sqrt{d}]$ i.e. all denominators of rational coefficients in
$\alpha,\beta$ divide $k$. Then $kI$ is an ideal and $kI\subseteq
\ZZ [\sqrt{d}] \subseteq \co$ i.e. $kI$ is an integral ideal.
$\qed$

Recall that $a\ZZ+\frac{b}{2}\ZZ = a\ZZ +\frac{b'}{2}\ZZ$ for any
$b' \equiv b \mod 2a$ so we can adjust the value of $b$ to lie in
any desired interval of length $2a$. We will make use of the
following basic choice:\\ For $a,b\in\ZZ$ with $a\neq 0$ let $\tau
(b,a)$ be the unique integer $\tau$ such that $\tau \equiv b \mod
2a$ and \\ $-a < \tau \leq a$ if $a>\sqrt{D}$,\\ $\sqrt{D}-2a
<\tau \leq \sqrt{D}$ if $a<\sqrt{D}$.

\begin{proposition}\label{num19} A subset $I\subseteq \numd$ is an
integral ideal of $\co$ {\bf iff} $I$ can be written as \beq
\label{rep1} I=k\left( a\ZZ+\frac{b+\sqrt{D}}{2}\ZZ\right) \eeq
where $a,b,k
\in \ZZ$ with $a>0$, $k>0$, $b=\tau (b,a)$ and $4a$ divides $b^2-D$. \\
Furthermore this presentation of the ideal $I$ as the triple of
integers $(a,b,k)$ is unique: $ak$ is the least positive rational
in $I$, $k/2$ is the least positive coefficient of $\sqrt{D}$ of
any member of $I$ and $b=\tau(b,a)$ uniquely determines $b$. Also
$\cn(I)=k^2a$.
\end{proposition}

\noindent {\bf Proof}\,\, ($\Rightarrow$) Suppose that $I$ is an
integral ideal of $\co$. By proposition \ref{num17} we have
\[ I=a'\ZZ+\beta\ZZ \hspace{5mm} \mbox{where $a'\in\ZZ$ is the least
positive integer in $I$.} \] $\beta \in I\subseteq \co$ so we can
write $\beta=m_1+\frac{D+\sqrt{D}}{2}m_2 = \frac{b'+k\sqrt{D}}{2}$
for integers $b',k$. Since $\beta\ZZ = -\beta\ZZ$ we can assume
that $k>0$. Now $a'\in I$ and $\frac{D+\sqrt{D}}{2}\in \co$ so
their product is in $I$ so there exist integers $m_1,m_2$ with
\[
a' \left( \frac{D+\sqrt{D}}{2}\right)
=m_1a'+m_2\beta=m_1a'+m_2(\frac{b'+k\sqrt{D}}{2}).\] Equating
coefficients of $\sqrt{D}$ gives $a'=m_2k$ so $k$ divides $a'$.
Also $\frac{b'+k\sqrt{D}}{2}\in I$ so similarly, there exist
integers $n_1,n_2$ with \[
(\frac{b'+k\sqrt{D}}{2})(\frac{D+\sqrt{D}}{2})=
n_1a'+n_2(\frac{b'+k\sqrt{D}}{2}). \] Equating coefficients of
$\sqrt{D}$ then shows that $k$ divides $b'$. Writing $a'=ka$ and
$b'=kb$ we get $I=k(a\ZZ+\frac{b+\sqrt{D}}{2}\ZZ)$. In this
expression $k/2$ is uniquely determined as the least positive
coefficient of $\sqrt{D}$ in any member of $I$. Also $ka$ is
uniquely determined as the least positive rational in $I$ so $a$
is unique too. The condition $b=\tau (b,a)$ then uniquely fixes
the value of $b$. Finally we show that $4a$ divides $D-b^2$. Since
$k(b+\sqrt{D})/2\in I$ and $(D+\sqrt{D})/2\in \co$ we have \[
\left( \frac{b+\sqrt{D}}{2}\right) \left( \frac{D+\sqrt{D}}{2}
\right) = k\frac{(b+1)D+(b+D)\sqrt{D}}{4} \in I. \] Hence it must
be of the form $k(xa+y\frac{b+\sqrt{D}}{2})$ for $x,y\in \ZZ$.
Thus $(b+D)=2y$ and $D(b+1)=4ax+2yb=4ax+b^2+bD$ giving $4ax=D-b^2$
i.e. $4a$ divides $D-b^2$.\\ ($\Leftarrow$) Conversely suppose
that $I$ has the form given in eq. (\ref{rep1}) with $a,b,k\in
\ZZ$ satisfying the given conditions. We show that $I$ is then an
ideal of $\co$.  Now $I$ is clearly closed under $\ZZ$-linear
combinations of its members so it remains to show that $\co\cdot
I=I$. Since $\co$ has integral basis $\{
1,\omega=\frac{D+\sqrt{D}}{2} \}$ it then suffices to show that
$1I\subseteq I$ and $\omega I\subseteq I$. The first is obvious
and for the second it suffices to show $ka\omega \in I$ and
$k(\frac{b+\sqrt{D}}{2})\omega\in I$. Now $ka\omega\in I$ iff
\[ ka\omega = k\left(\frac{aD}{2}+\frac{a}{2}\sqrt{D}\right)
 = m_1ka+m_2k
\left(\frac{b+\sqrt{D}}{2}\right) \] has a solution $m_1,m_2$ in
{\em integers}. Equating coefficients gives $m_1a+m_2b/2=aD/2$ and
$m_2/2=a/2$ so $m_2=a$ and $m_1=(D+b)/2$. But $b^2 =D+4ac$ so
$b^2\equiv D\mod 2$ so $b\equiv D\mod 2$ (as $b$ and $b^2$ always
have the same parity). Hence $m_1, m_2$ are integers as
required.\\ A similar calculation for
$k(\frac{b+\sqrt{D}}{2})\omega$ gives $m_2=(D+b)/2$ which is an
integer and $m_1=\frac{D-b^2}{4a}$ which is an integer by
hypothesis.

Finally $\cn(I)=k^2a$ follows by direct calculation with the
integral basis  $ \{ ka, k\frac{b+\sqrt{D}}{2}\}$. $\qed$

\noindent {\bf Remark} A principal ideal $I=\alpha\co$ can be
given either by giving a value of $\alpha$ or by giving the
parameters $a,b,k\in\ZZ$. Although there is a $O(\poly (\log
\alpha, \log D))$ time algorithm for translating $\alpha$ into
$(a,b,k)$ (cf. proposition \ref{numtransl} below) the reverse
translation appears to be a hard computational task (classically)
-- the best known classical algorithm has running time
$O(e^{\sqrt{\log d}})$. Thus this interconversion corresponds to a
one-way function which forms the basis of the Buchmann-Williams
cryptosystem \cite{BW89} for key exchange (which can be broken by
an extension \cite{hallgren} of Hallgren's algorithm).

\begin{proposition} \label{numtransl}  Let
$\alpha=\frac{x+y\sqrt{D}}{2} \in \co$ with $x,y\in \ZZ$. Let
$k=\gcd (y, (x+yD)/2)$ and $u,v\in \ZZ$ such that \[ uy+v(x+yD)/2
=k \] (which are guaranteed to exist by Euclid's algorithm). Then
\[ \alpha\co= k(a\ZZ+\frac{b+\sqrt{D}}{2}\ZZ ) \] with $k$ as above,
$a=|\alpha \overline{\alpha}|/k^2$ and
$b=\tau((ux+y(x+yD)/2)/k,a)$. Hence the parameters $(a,b,k)$ of
$\alpha\co$ can be computed in $\poly(\log |x|, \log |y|, \log D)$
time. \end{proposition}

\noindent {\bf Proof}\,\, Since $\co=\ZZ+\frac{D+\sqrt{D}}{2}\ZZ$,
it follows that any $\alpha\in \co$ may be written as
$\alpha=x'+y'\frac{D+\sqrt{D}}{2}$ for $x',y'\in \ZZ$ so
$\alpha=\frac{x+y\sqrt{D}}{2}$ with $x,y\in \ZZ$ and $(x+yD)/2 \in
\ZZ$. Also then, $\alpha\co$ is generated (via $\ZZ$-linear
combinations) by $(x+y\sqrt{D})/2$ and
$(x+y\sqrt{D})(D+\sqrt{D})/4= ((x+y)D+(x+yD)\sqrt{D})/4$. Since
$k$ is the smallest positive coefficient of $\sqrt{D}/2$ we get
$k=\gcd (y, (x+yD)/2)$. Also $\cn (\alpha \co)=k^2a =
|\alpha\overline{\alpha}|$ giving the claimed formula for $a$.

The only elements of $\alpha\co$ with $\sqrt{D}$ coefficient $k/2$
have the form $k(aw+\frac{b+\sqrt{D}}{2}1)=k(b'+\sqrt{D})/2$ where
$b'=b+2wa$ and $w\in \ZZ$, so then $b=\tau(b',a)$. Now if
$k=uy+v(x+yD)/2$ for $u,v\in \ZZ$ then
$u\frac{x+y\sqrt{D}}{2}+v((x+y)D+(x+yD)\sqrt{D})/4$ is in
$\alpha\co$ and has the form $k(b'+\sqrt{D})/2$ where
$b'=(ux+v(x+y)/2)/k$. Then $b=\tau (b',a)$ gives the claimed
formula for $b$.

Finally in $k=uy+v(x+yD)/2$, $u$ and $v$ might conceivably be very
large but the choice is not unique -- we have the freedoms
$u\rightarrow u-s(x+yD)/2$ and $v\rightarrow v+sy$ for any $s\in
\ZZ$. Hence we can take $v<y$ and then $u=(k=v(x+y)D/2)/y$ is also
suitably small making the whole computation of $a,b,k$ performable
in $\poly (\log |x|, \log |y|, \log D )$ time. $\qed$

Using proposition \ref{num18} with proposition \ref{num19} we get
a corresponding unique presentation of a fractional ideal as \beq
\label{frid} I=\frac{k}{l}\left(
a\ZZ+\frac{b+\sqrt{D}}{2}\ZZ\right) \eeq with $l\in\NN$ being the
smallest such integer and $a,b,k\in \ZZ$ satisfying the same
conditions as above. When the parameters $a,b,k,l$ satisfy these
conditions, making them unique, we say that $I$ is in standard
form.

%An integral ideal is called {\bf primitive} if $k=1$. From eq.
%\ref{frid}) it follows that if $I$ is any fractional ideal then
%there is $r\in\QQ$ such that $rI$ is a primitive integral ideal.
%(We take $r=d/k$).

\section{Reduced ideals and the reduction operator
$\rho$}\label{redsect}

\subsection{Reduced ideals}\label{redsect1}

It will be helpful to introduce a geometrical picture of a
fractional ideal $I$ as a two dimensional lattice embedded in
$\RR^2$: if $\alpha\in I$ we map it to the point $\hat{\alpha}=
(\alpha, \overline{\alpha})\in \RR^2$ (recalling that for $\alpha
= p+q\sqrt{d}$ we define $\overline{\alpha}= p-q\sqrt{d}$).

\begin{definition}\label{min} A {\bf minimum} of $I$ is an element
$\alpha \in I$ such that $\alpha >0$ and there is no nonzero
$\beta \in I$ with $|\beta|<|\alpha|$ and
$|\overline{\beta}|<|\overline{\alpha}|$ i.e. $\hat{\alpha}\in
\RR^2$ lies in the right half plane and the rectangle defined by
$\hat{\alpha}$ (having corners at the four points $(\pm \alpha,
\pm \overline{\alpha})$ contains no lattice points inside it
(except for $(0,0)$).\\ A fractional ideal $I$ is called {\bf
reduced} if $1\in I$ and $1$ is a minimum of $I$. \end{definition}

\begin{proposition} \label{num101} If $I$ is reduced then its standard form is
\[ I=
\ZZ+\frac{b+\sqrt{D}}{2a}\ZZ \] i.e. in eq. (\ref{frid}) we have
$k=1$ and $l=a$. \end{proposition}

\noindent {\bf Proof}\,\, Since $1\in I$ we have
$1=\frac{k}{l}(xa+yb+y\sqrt{D})$ with $x,y \in \ZZ$. Hence $y=0$
and $kxa/l=1$. But $1\in I$ is also a minimum so $x=1$ (because if
$x>1$ then $\beta =k(x-1)a/l <1$ is an element of $I$ with
$0<\beta<1$ and $0<\overline{\beta}<1$). Thus $ka=l$ so $k$
divides $l$ and the minimality of $l$ in eq. (\ref{frid}) implies
$k=1$. $\qed$

\begin{proposition} \label{num102} Let \[ I=\redid \] be a reduced
ideal in standard form. Then $a< \sqrt{D}$ and $|b|<\sqrt{D}$.
Hence the number of reduced ideals is finite. \end{proposition}

\noindent {\bf Proof}\,\, From the definition of $\tau$ and
$b=\tau (b,a)$ we have $|b|<\sqrt{D}$ if $a<\sqrt{D}$. To show
$a<\sqrt{D}$ suppose on the contrary that $a>\sqrt{D}$. Then from
the definition of $\tau$ we have $|b|<a$ so
$|\frac{b+\sqrt{D}}{2a}|<1$ and $|\frac{b-\sqrt{D}}{2a}|<1$
contradicting the fact that 1 is a minimum (since
$\frac{b+\sqrt{D}}{2a}\in I$). Hence $a<\sqrt{D}$. $\qed$

\begin{proposition}\label{num103} Suppose a fractional ideal has
the standard form \[ I= \redid . \] Then $I$ is reduced {\bf iff}
$b\geq 0$ and $b+\sqrt{D}>2a$. \end{proposition}

\noindent {\bf Proof}\,\, ($\Rightarrow$) Assume $I$ is reduced.
By proposition \ref{num102} $a<\sqrt{D}$ and so $b=\tau(b,a)$ lies
in the range $\sqrt{D}-2a<b<\sqrt{D}$. If $b<0$ then $|b|=-b
<2a-\sqrt{D}$ i.e. $|b|+\sqrt{D}<2a$ which contradicts the
minimality of 1 (as $\max (|\frac{b+\sqrt{D}}{2a}|,
|\frac{b-\sqrt{D}}{2a}|)=\frac{|b|+\sqrt{D}}{2a}$). Hence $b\geq
0$ and if $\frac{b+\sqrt{D}}{2a}<1$ then $\max
(|\frac{b+\sqrt{D}}{2a}|,
|\frac{b-\sqrt{D}}{2a}|)=\frac{|b|+\sqrt{D}}{2a}=
\frac{b+\sqrt{D}}{2a}<1$ again contradicting the minimality of 1.
Hence $b+\sqrt{D}>2a$.\\ ($\Leftarrow$) Conversely assume that
$b\geq 0$ and $b+\sqrt{D}>2a$. Let $H(x,y)=\max
(|2xa+y(b+\sqrt{D})|, |2xa +y(b-\sqrt{D})|)$. Then $I$ is reduced
iff 1 is a minimum of $I$ iff $H(x,y)>2a$ for all $x,y\in \ZZ$
with $(x,y)\neq (0,0)$. Since $H(x,y)=H(-x,-y)$ it suffices to
consider the two sectors $x\geq 0,y\geq0$ and $x>0,y\leq 0$.\\ If
$x\geq 0,y\geq 0$ and $(x,y)\neq (0,0)$ then $b+\sqrt{D}>2a$ gives
$|2xa+y(b+\sqrt{D})|\geq 2a(x+y)\geq 2a$ so $H(x,y)\geq 2a$ in
this sector. \\ For $x>0,y\leq 0$ we first show that
$b-\sqrt{D}<0$. Indeed $b+\sqrt{D}>2a$ gives $a<\sqrt{D}+(b-a)$.
If $a>\sqrt{D}$ then the definition of $\tau$ gives $b<a$ and the
contradiction $a<\sqrt{D}$. Hence $a<\sqrt{D}$ and the definition
of $\tau$ again gives $b<\sqrt{D}$ i.e. $b-\sqrt{D}<0$. Then
$|2xa+y(b-\sqrt{D})|=2ax+|y||b-\sqrt{D}|\geq 2xa$. Since $x>0$ we
get $H(x,y)\geq 2a$. $\qed$

\begin{corollary} \label{num104} The fractional ideal $I=\redid$
is reduced if $a\leq \sqrt{D}/2$. \end{corollary}

\noindent {\bf Proof}\,\, If $a\leq \sqrt{D}/2$ then the
definition of $\tau$ gives $b>0$ and then also $b+\sqrt{D}\geq
b+2a
>2a$ so by proposition \ref{num103} $I$ is reduced. $\qed$

\subsection{The reduction operator $\rho$}\label{redsect2}

For a fractional (not necessarily reduced) ideal of the form \[
I=\redid
\] we introduce the notation
\begin{equation}\label{gammadef} \gamma(I)=\frac{b+\sqrt{D}}{2a}
\end{equation} and define the {\bf reduction operator} $\rho$
mapping (principal) ideals to (principal) ideals by \[
\rho(I)=\frac{1}{\gamma(I)} I = \frac{2a}{b+\sqrt{D}}\ZZ+\ZZ. \]
Since $4a$ divides $b^2-D$ (cf. proposition \ref{num19}) we
introduce the integer $c=|D-b^2|/(4a)$ and then \[
\frac{2a}{b+\sqrt{D}}=\frac{2a(b-\sqrt{D})}{b^2-D}=
\pm\frac{b-\sqrt{D}}{2c}.
\] Thus (as $\ZZ=\pm\ZZ$) \[ \frac{2a}{b+\sqrt{D}}\ZZ+\ZZ= \frac{-b+\sqrt{D}}{2c}\ZZ
+\ZZ = \ZZ+\frac{\tau(-b,c)+\sqrt{D}}{2c} \ZZ .\] Hence the
standard parameters $a',b'$ of $\rho(I)= \ZZ
+\frac{b'+\sqrt{D}}{2a'}\ZZ$ are \begin{equation}\label{redparm}
b'=\tau(-b,c) \hspace{1cm} a'=c=\frac{|D-b^2|}{4a}. \end{equation}

\noindent {\bf Remark}\,\, Many treatments of this subject use a
slightly more general notion of reduction. Recall that a general
fractional ideal has the standard form
\[ I=\frac{k}{l}\left( a\ZZ+\frac{b+\sqrt{D}}{2}\ZZ\right) =
\frac{ak}{l} \left( \redid \right) \] with $a,b,k,l$ uniquely
determined. Write $\gamma (I)=\frac{b+\sqrt{D}}{2a}$.  Then in the
more general formalism the reduction operator is defined by
$\rho(I)=\frac{1}{\gamma(I)}I$ and $I$ is said to be reduced if
$ka/l$ (the smallest positive rational in $I$) is a minimum. Thus
$I$ is reduced iff $\ZZ+\gamma(I)\ZZ$ is reduced in our restricted
sense (i.e. for the restricted sense that we're using, we also
require 1 to be the smallest positive rational in $I$).

\begin{proposition}\label{num105}
Let $I=\redid$ be a (not necessarily reduced) fractional ideal.
Let $I_0=I$ and $I_i=\rho(I_{i-1})=
\ZZ+\frac{b_i+\sqrt{D}}{2a_i}\ZZ$. If $I_i$ is not reduced then
$a_i <a_{i-1}/2$ and thus $I_i$ is reduced for some $i\leq \lceil
\log_2 (a/\sqrt{D})\rceil +1$. Let $i_{\rm red}$ be the minimal
such $i$. Then $\alpha=\prod_{j=1}^{i_{\rm red}-1} \gamma(I_j)$ is
a minimum in $I$ and $I_{\rm red}= I_{i_{\rm
red}}=\frac{1}{\alpha}I$.
\end{proposition}

\noindent {\bf Proof}\,\, If $a_{i-1}>\sqrt{D}$ then
$|b_{i-1}|<a_{i-1}$ so \[ a_i=\frac{|b_{i-1}^2-D|}{4a_{i-1}} <
\frac{a_{i-1}^2+D}{4a_{i-1}}<\frac{a_{i-1}}{2}. \] Hence
$a_i<\sqrt{D}$ for some $i\leq \lceil \log (a/\sqrt{D}) \rceil$.\\
Now assume that $a_i<\sqrt{D}$ and consider the two cases
$a_{i+1}\geq a_i$ and $a_{i+1}<a_i$. In the first case, we have
$|b_i^2-D|/(4a_i)\geq a_i$ and also $\sqrt{D}-2a_i <b_i<\sqrt{D}$
so $D-b_i^2\geq 4a_i^2$ giving $\sqrt{D}+|b_i|>2a_i$. Next we show
that $b_i\geq 0$, so by proposition \ref{num103}, $I$ will be
reduced. If $b_i<0$ then $\sqrt{D}-2a_i<b_i<0$ gives
$|b_i|<|\sqrt{D}-2a_i|=2a_i-\sqrt{D}$ in contradiction to
$\sqrt{D}+|b_i|>2a_i$. Hence $b_i\geq 0$.\\ In the second case: if
$a_{i+1}<a_i$ then $a_{i+1}^2 <
a_ia_{i+1}=a_i|\frac{b_i^2-D}{4a_i}| = \frac{D-b_i^2}{4}$. Thus
$a_{i+1}<\sqrt{D}/2$ and corollary \ref{num104} implies that
$I_{i+1}$ is reduced.\\ Finally we prove the statement about
$\alpha$. Clearly $I_{\rm red}=I/\alpha$ and since $I_{\rm red}$
is reduced it follows that $\alpha$ must be a minimum in $I$. To
see this think of the geometrical picture: we have $I=\alpha
I_{\rm red}$ so the lattice for $I$ is obtained from the lattice
for $I_{\rm red}$ by rescaling by $\alpha$ and $\overline{\alpha}$
in the $x$ and $y$ directions respectively. Hence the interior of
the $\hat{\alpha}$ rectangle for $I$ corresponds to the interior
of the $\hat{1}$ rectangle for $I_{\rm red}$ which is empty except
for the point $(0,0)$. $\qed$

The {\bf right neighbour} of a minimum $\alpha \in I$ is the
uniquely determined minimum $\beta_R\in I$ with least size
$\beta_R>\alpha$. The {\bf left neighbour} of a minimum $\alpha\in
I$ is the uniquely determined minimum $\beta_L\in I$ with least
conjugate size $|\overline{\beta_L}|>|\overline{\alpha}|$.

\begin{proposition}\label{num106} Let $\alpha \in \numd$ with $\alpha >0$.
Then for
any fractional ideal $I$ the mapping $I\rightarrow \alpha I$ is a
bijection mapping minima to minima and left (resp. right)
neighbours to left (resp. right) neighbours. \end{proposition}

\noindent {\bf Proof}\,\, This is immediate from the geometrical
picture of multiplication by $\alpha$ as just rescaling the
lattice by $\alpha$ and $\overline{\alpha}$ in the $x$ and $y$
directions respectively. $\qed$

\begin{proposition} \label{num107} If $I=\ZZ +\gamma(I)\ZZ$ is
reduced then $\gamma(I)>1$ and $-1<\overline{\gamma(I)}<0$.
\end{proposition}

\noindent {\bf Proof}\,\, By proposition \ref{num103} if $I$ is
reduced then $\gamma(I)=(b+\sqrt{D})/(2a)>1$. Also by proposition
\ref{num102} $a<\sqrt{D}$ so $\sqrt{D}-2a<b<\sqrt{D}$ so
$-1<(b-\sqrt{D})/(2a)<0$. $\qed$

\begin{proposition}\label{num108} If $I=\ZZ+\gamma(I)\ZZ$ is
reduced then $\gamma(I)$ is a minimum of $I$ and $\rho(I)$ is
again reduced.\end{proposition}

\noindent {\bf Proof}\,\, Write $\gamma = \gamma(I)$. To see that
it is a minimum of $I$ we need to show that if $x+y\gamma$ (with
$x,y\in \ZZ$) has $|x+y\gamma|<|\gamma|$ and
$|x+y\overline{\gamma}|<|\overline{\gamma}|$ then $x=y=0$. Without
loss of generality we may assume $x\geq 0$. Suppose the above
conditions hold for $x,y\in \ZZ$. Since $x\geq 0$, $\gamma >1$ and
$-\gamma < x+y\gamma <\gamma$ we see that $y\leq 0$. Write
$\xi=|\overline{\gamma}|$ so $0<\xi<1$ and
$|x+y\overline{\gamma}|<|\overline{\gamma}|$ gives
$-\xi<x-y\xi<\xi$. Since $x\geq 0$ we get $-y\xi<\xi$ so $y>-1$.
But we had $y\leq 0$ so $y=0$ is the only possibility. Then
$|x|<|\xi|<1$ gives $x=0$. Thus $\gamma$ is a minimum of $I$ and
by proposition \ref{num106} $\rho(I)$ is again a reduced ideal.
$\qed$

\begin{proposition}\label{num109} Let $I=\ZZ+\gamma(I)\ZZ$ be a
reduced ideal. Then the minimum $\gamma(I)\in I$ is the right
neighbour of 1 in $I$.
\end{proposition}

\noindent {\bf Proof}\,\, Suppose that $\alpha =x+y\gamma(I)$ is
the right neighbour of 1. We first show that $y>0$. If $y=0$ then
$\alpha=x$ so $x>1$ i.e. $\alpha\geq 2$,
$\alpha=\overline{\alpha}$ and the $\hat{\alpha}$ rectangle
contains $(1,1)$ i.e. $\alpha$ cannot be a minimum.\\ If $y<0$ put
$y=-k$ for $k>0$. Then $\alpha=x-k\gamma>1$ and $\gamma>1$ gives
$x>1+k\gamma >1+k$. Then
$\overline{\alpha}=x-k\overline{\gamma}>1+k-k\overline{\gamma}>1+k$
(as $\overline{\gamma}<0$). Hence again the $\hat{\alpha}$
rectangle contains $(1,1)$ so $\alpha$ cannot be a minimum. Thus
$y>0$. Next we show that $x\geq 0$. If $x<0$ put $x=-k$ for $k>0$.
Then $\alpha=-k+y\gamma>1$ and
$\overline{\alpha}=-k+\overline{\gamma}<-k\leq -1$ i.e.
$\overline{\alpha}<-1$ and the $\hat{\alpha}$ rectangle contains
$(1,1)$ i.e. $\alpha$ cannot be a minimum. Hence $\alpha
=x+y\gamma$ with $x\geq 0$ and $y>0$ are the only possibilities
and the smallest such $\alpha>1$ is $\alpha =\gamma$ which
actually is a minimum by proposition \ref{num108}. $\qed$

\subsection{The principal cycle of reduced ideals}\label{redsect3}

Since $\co=\ZZ
+\frac{D+\sqrt{D}}{2}\ZZ=\ZZ+\frac{\tau(D,2)+\sqrt{D}}{2}\ZZ$ we
see from proposition \ref{num103} that $\co$ is a reduced
principal ideal of $\co$. Thus $\alpha_0=1\in \co$ is a minimum
and for $i\in \ZZ$ let $\alpha_{i-1}$ be the left neighbour and
$\alpha_{i+1}$ the right neighbour of the minimum $\alpha_i \in
\co$. Also let $J_i=\frac{1}{\alpha_i}\co = \ZZ +\gamma_i\ZZ$.
(Note that since $\alpha_i$ is a minimum of $\co$, proposition
\ref{num106} shows that $1\in J_i$ is a minimum so $J_i$ is
reduced and hence by proposition \ref{num101} has the form
$\ZZ+\gamma_i\ZZ$.)

Since $\alpha_{i+1}$ is the right neighbour of $\alpha_i$ in $\co$
it follows that $\alpha_{i+1}/\alpha_i$ is the right neighbour of
1 in $\frac{1}{\alpha_i}\co$ and proposition \ref{num109} gives
$\alpha_{i+1}/\alpha_i=\gamma_i$ i.e.
\[ \alpha_{i+1}=\alpha_i \gamma_i \hspace{1cm} \mbox{and
$J_{i+1}=\rho (J_i)$.} \] {\bf Remark}\,\, (Geometrical picture of
the sequence of minima) If we plot the points $\hat{\alpha}_i\in
\RR^2$ they all lie in the right half of the plane ($\alpha_i>0$)
on a pair of hyperbola--like curves (like $y=\pm 1/x$) alternating
with $\pm y$ values as $i$ varies. To see this recall that
$\alpha_{i+1}>\alpha_i$ (by definition) and so
$|\overline{\alpha}_{i+1}|<|\overline{\alpha}_i|$ (since
$|\overline{\alpha}_{i+1}|>|\overline{\alpha}_i|$ would imply that
$\hat{\alpha}_i$ is inside the $\hat{\alpha}_{i+1}$ rectangle)
i.e. the sequence $\{ |\overline{\alpha}_i|\}$ is monotonically
decreasing with increasing $i$. To see that $\overline{\alpha}_i$
and $\overline{\alpha}_{i+1}$ have opposite signs consider
$J_i=\co/\alpha_i = \ZZ+\gamma_i\ZZ$. As noted above we have
$\alpha_{i+1}/\alpha_i=\gamma_i$ and proposition \ref{num107}
gives $-1<\overline{\gamma}_i<0$ so
$\overline{\alpha}_{i+1}/\overline{\alpha}_i <0$.

\begin{proposition}\label{num110} For all $i\in \ZZ$ \[ \ln
\alpha_{i+1}-\ln \alpha_i \leq \frac{1}{2}\ln D \]
\end{proposition}

\noindent {\bf Proof}\,\, We have $\alpha_{i+1}/\alpha_i=\gamma_i$
where $J_i=\ZZ+\gamma_i\ZZ$ is a reduced ideal and
$\gamma_i=\frac{b_i+\sqrt{D}}{2a_i}$. Since $J_i$ is reduced,
proposition \ref{num102} gives $b_i<\sqrt{D}$ so $\gamma_i <
\frac{2\sqrt{D}}{2a_i}<\sqrt{D}$ giving $\ln (
\alpha_{i+1}/\alpha_i)\leq \ln \sqrt{D}$. $\qed$

\begin{proposition}\label{num111} For all $i\in\ZZ$ \[ \ln
\alpha_{i+1}-\ln \alpha_i \geq \ln \left( 1+\frac{3}{16D}\right)
\geq \frac{3}{32D}. \] {\em (Remark: In \cite{will02} (at eq.
(5.4)) a stronger lower bound of $\ln (1+1/\sqrt{D})$ is claimed
but the above will suffice for our purposes.)}\end{proposition}

\noindent {\bf Proof}\,\, As in the proof of proposition
\ref{num110} we have
\[
\frac{\alpha_{i+1}}{\alpha_i}=\gamma_i=\frac{b_i+\sqrt{D}}{2a_i}>1.
\] Omitting the subscripts $i$ we have
$\frac{b+\sqrt{D}}{2a}=1+\frac{b+\sqrt{D}-2a}{2a}$. Let $\epsilon
= \sqrt{D} - \lfloor \sqrt{D} \rfloor$ and
$\xi=\frac{b+\sqrt{D}-2a}{2a}= \frac{\sqrt{D}-(2a-b)}{2a}$. Then
since $0<a<\sqrt{D}$ and $\xi>0$ with $2a-b\in \ZZ$ we have \[ \xi
> \frac{\epsilon}{2\sqrt{D}}. \]
Write $K=\lfloor \sqrt{D}\rfloor$ and $D=K^2+L$ with
$1<L<2K+1<K^2$ so $\sqrt{D}=K\sqrt{1+L/K^2}$. We apply the
binomial inequality \[ 1+x/2 > \sqrt{1+x}>1+x/2-x^2/8 \] and for
$0<x<1$ we have $x^2<x$ so $\sqrt{1+x}>1+3x/8$. Hence \[
\sqrt{D}>K+\frac{3L}{8K}>K+\frac{3}{8K}>K+\frac{3}{8\sqrt{D}} \]
as $K<\sqrt{D}$. Thus $\epsilon > \frac{3}{8\sqrt{D}}$ and
$\xi>\frac{3}{16D}$ so $\alpha_{i+1}/\alpha_i>1+\frac{3}{16D}$.
Finally using (for $0<x<1$) \[ \ln (1+x) >x-x^2/2 >x-x/2 =x/2 \]
we get $\ln (\alpha_{i+1}/\alpha_i) >\ln (1+\frac{3}{16D}
>\frac{3}{32D}$. $\qed$

Thus we have upper and lower bounds on the separation between
consecutive $\ln \alpha_i$ values. However the lower bound is
exponentially small (in $\log D$) and we next give a constant
lower bound for the separation between every second member of the
sequence.

\begin{proposition}\label{num112} For all $i\in \ZZ$ \[ \ln
\alpha_{i+1}-\ln \alpha_{i-1} \geq \ln2 .\] \end{proposition}

\noindent {\bf Proof}\,\, We saw previously that
$\overline{\alpha}_{i-1}$ and $\overline{\alpha}_{i+1}$ have the
same sign and the sequence $|\overline{\alpha}_i|$ is strictly
decreasing with $i$. Hence
$|\overline{\alpha}_{i+1}-\overline{\alpha}_{i-1}|<
|\overline{\alpha}_{i-1}|$. Now if $\alpha_{i+1}/\alpha_{i-1} <2$
the we would have $\alpha_{i-1}<\alpha_{i+1}<2\alpha_{i-1}$ so
$0<\alpha_{i+1}-\alpha_{i-1}<\alpha_{i-1}$ i.e. if we write $\beta
= \alpha_{i+1}-\alpha_{i-1}$ then $\hat{\beta}$ lies inside the
$\hat{\alpha}_{i-1}$-rectangle. Since $\alpha_{i\pm 1}\in \co$ we
have $\beta \in \co$ and this contradicts the minimality of
$\alpha_{i-1}$. Hence $\alpha_{i+1}/\alpha_{i-1}\geq 2$. $\qed$

\begin{proposition} \label{num113} The sequence $\{ \alpha_i \}$
contains all the minima of $\co$.\end{proposition}

\noindent {\bf Proof}\,\, Let $\alpha\in \co$ be any minimum. From
propositions \ref{num110} and \ref{num112} we see that
$\lim_{i\rightarrow \pm\infty} \alpha_i = \pm\infty$ so there must
be an $i\in\ZZ$ with $\alpha_i\leq \alpha <\alpha_{i+1}$. We claim
$\alpha = \alpha_i$. Otherwise $\alpha>\alpha_i$ contradicting the
definition that $\alpha_{i+1}$ is the minimum with least size
greater than $\alpha_i$. $\qed$

\begin{theorem}\label{num114} {\bf (The principal cycle of reduced ideals).} \\ (a)
The sequence $\{J_i\}_{i\in\ZZ}$ is periodic i.e. there is (a
smallest) $k_0\in \NN$ such that $J_i=J_j$ {\bf iff} $i \equiv j
\mod k_0$ for all $i,j\in \ZZ$. The repeating segment $\{ J_0=\co,
J_1, \ldots J_{k_0-1} \}$ of principal reduced ideals is called
the {\bf principal cycle}.\\ (b) Let $\epsilon =
\alpha_{k_0}/\alpha_0 = \alpha_{k_0}$. Then $\epsilon =
\epsilon_0$, the fundamental unit of $\co$. \\ (c) Let $I$ be any
reduced principal fractional ideal. Then $I$ is in the principal
cycle.
\end{theorem}

\noindent {\bf Proof}\,\, (a) By proposition \ref{num102} the
sequence $\{ J_i\}_{i\in\ZZ}$ contains only finitely many
different ideals. Hence for some $i,k\in\ZZ$ we have $J_i=J_{i+k}$
so $\alpha_i \co = \alpha_{i+k}\co$. Write
$\eta=\alpha_{i+k}/\alpha_i$. Since $\alpha_{i+k}=\eta\alpha_i$ it
follows from proposition \ref{num106} that
$\alpha_{s+k}=\eta\alpha_s$ for any $s\in\ZZ$ and
$J_s=J_{s+k}=J_{s+lk}$ for all $l,s\in\ZZ$. Set $s=0$ and choose
the minimal $k_0$ such that $\co=J_{k_0}=\alpha_{k_0}\co$. Write
$\epsilon = \alpha_{k_0}$ which is necessarily a unit of $\co$ by
proposition \ref{num13}. Then $J_i$ for $0\leq i<k_0$ are pairwise
distinct and \begin{equation}\label{same} J_s=J_t \hspace{4mm}
\mbox{iff $s\equiv t \mod k_0$ for $s,t\in\ZZ$ iff
$\alpha_s=\alpha_t \epsilon^l$ for some $l\in\ZZ$} \end{equation}
(b) To see that $\epsilon = \alpha_{k_0}$ is the fundamental unit
$\epsilon_0$ of $\co$ let $\eta$ be any unit with $\eta>0$. Then
$\eta$ is a minimum of $\co$ (because if $\xi \in \co$ has
$|\xi|<|\eta|$ and $|\overline{\xi}|<|\overline{\eta}|$ then
$\xi/\eta \in \co$ has $|\xi/\eta|<1$ and $|\overline{\xi/\eta}| =
|\overline{\xi}/ \overline{\eta}|<1$ contradicting the minimality
of 1 in $\co$). Hence by proposition \ref{num113} $\eta\co=J_k$
for some $k\in\ZZ$. But $\eta\co=\co$ if $\eta $ is a unit so eq.
(\ref{same}) gives $\eta=\epsilon^l$ for some $l\in\ZZ$. Hence
$\epsilon = \epsilon_0$ the fundamental unit. \\ (c) Since $I$ is
principal we can write $I=\frac{1}{\alpha}\co$. Hence $\co=\alpha
I$ and since 1 is a minimum of $I$ proposition \ref{num106} shows
that $\alpha$ is a minimum of $\co$. Then proposition \ref{num113}
shows that $I=J_k$ for some $k$. $\qed$

\begin{proposition} \label{num115} The length $k_0$ of the
principal cycle satisfies \[ \frac{2R}{\ln D} \leq k_0 \leq
\frac{2R}{\ln 2}. \] \end{proposition}

\noindent {\bf Proof}\,\, We have $R=\ln \epsilon_0= \ln
\alpha_{k_0}= \ln \alpha_{k_0}-\ln \alpha_0$ (as $\alpha_0=1$).
Then proposition \ref{num112} gives $\frac{k_0}{2}\ln 2 \leq R$
i.e $k_0\leq 2R/\ln 2$. Similarly proposition \ref{num110} gives
$k_0 (\frac{1}{2}\ln D) \geq R$. $\qed$

\subsection{The inverse of the reduction operator}\label{redsect4}

Recall that if we apply the reduction operator $\rho$ sufficiently
many times to $I=\ZZ+\frac{b+\sqrt{D}}{2a}\ZZ$ we will always
eventually obtain a reduced ideal. As $a>0$ can be arbitrarily
large, there are infinitely many distinct ideals of this form. We
also know that there are only finitely many reduced ideals and if
$I$ is reduced then $\rho(I)$ is reduced too. Hence as a mapping
on general ideals $\rho$ cannot be one-to-one.

However if we restrict to the subset of reduced principal ideals
i.e. the principal cycle $\cp\ci_{\rm red}= \{ J_0=\co, J_1,
\ldots , J_{k_0-1}\}$ then we have $J_{i+1}=\rho(J_i)$ (cycling
with $\rho(J_{k_0-1})=J_0$) so $\rho$ is invertible and we now
develop an explicit expression for the inverse map $\rho^{-1}$.
Then together with the formula for the action of $\rho$ (in eq.
(\ref{redparm})) we will be able to step in either direction along
the principal cycle.

Let $I=\ZZ+\frac{b+\sqrt{D}}{2a}\ZZ=\ZZ+\gamma\ZZ$ be any reduced
ideal. We define the conjugated ideal \[ \sigma (I) =
\overline{I}=\ZZ+\frac{b-\sqrt{D}}{2a}\ZZ = \ZZ
+\frac{\tau(-b,a)+\sqrt{D}}{2a}\ZZ \] where the last expression
(using $\ZZ=-\ZZ$) is in standard form. In terms of the
geometrical picture of ideals, with $\alpha\in I$ embedded in
$\RR^2$ as $\hat{\alpha}=(\alpha,\overline{\alpha})$, we see that
the conjugation operation simply reflects the lattice in the
$45^\circ$
line $y=x$. Then the following facts are immediately evident: \\
(i) If $I$ is reduced then $\overline{I}$ is reduced (i.e. 1 stays
a minimum under conjugation). \\ (ii) If $\alpha$ is a minimum in
$I$ then $|\overline{\alpha}|$ is a minimum in $\overline{I}$. \\
(iii) If $\alpha$ is the left (right) neighbour of a minimum
$\beta$ in $I$ then $|\overline{\alpha}|$ is the right (left)
neighbour of the minimum $|\overline{\beta}|$ in $\overline{I}$.

Now write $b_*=\tau(-b,a)$. Then since $\delta=
\frac{\tau(-b,a)+\sqrt{D}}{2a}=\frac{b_*+\sqrt{D}}{2a}$ is the
right neighbour of 1 in the reduced ideal $\overline{I}$ (cf.
proposition \ref{num109}), we see that \[
|\overline{\delta}|=\left| \frac{b_*-\sqrt{D}}{2a}\right| =
\frac{\sqrt{D}-b_*}{2a} \] is the left neighbour of 1 in $I$.
Hence by the definition of $\rho (I)=\frac{1}{\gamma}I$ we get \[
\rho^{-1}(I)= \frac{1}{|\overline{\delta}|}I=
\frac{2a}{\sqrt{D}-b_*}\left( \ZZ +
\frac{b+\sqrt{D}}{2a}\ZZ\right). \] To express this in standard
form note that $-b$ and $b_*$ differ by a multiple of $2a$ and
$\ZZ=-\ZZ$ so we get $ \ZZ
+\frac{b+\sqrt{D}}{2a}\ZZ=\ZZ+\frac{b_*-\sqrt{D}}{2a}\ZZ$ so \[
\rho^{-1}(I)=\ZZ +\frac{2a}{\sqrt{D}-b_*}\ZZ = \ZZ
+\frac{b_*+\sqrt{D}}{2c_*}\ZZ \] where $c_*=(D-b_*^2)/(4a)$. Hence
if $\rho^{-1}(I)=\ZZ+\frac{b''+\sqrt{D}}{2a''}\ZZ$ is the standard
form, we have the explicit formulae: \beq \label{invparm}
b_*=\tau(-b,a) \hspace{1cm} a''=\frac{D-b_*^2}{4a} \hspace{1cm}
b''=\tau(b_*,a''). \eeq {\bf Remark}\,\, From the geometrical
picture we also see that $\rho^{-1}(I)=\sigma\rho\sigma (I)$ and
above we saw that $\sigma$ induces the mapping $a\rightarrow a$,
$b\rightarrow b_*$. These formulae together with eq.
(\ref{redparm}) may also be used to derive the above expressions
for $\rho^{-1}$. $\qed$

Since $0<a,b<\sqrt{D}$ for reduced ideals we see from eqs.
(\ref{redparm}) and (\ref{invparm}) that the action of $\rho$ and
$\rho^{-1}$ may be computed in $\poly \log D$ time.

\section{The distance function for ideals}\label{distance}

Let $I_1$ and $I_2$ be fractional (principal) ideals of $\co$
which are related by $I_1=\gamma I_2$ for some $\gamma \in \numd$.
Then the {\bf distance} $\delta (I_1, I_2)$ is defined by \beq
\label{dist}\delta (I_1, I_2)= \ln |\gamma| \mod R \eeq (recalling
that $R=\ln \epsilon_0$ where $\epsilon_0$ is the fundamental unit
of $\numd$). If $I_1 \neq \gamma I_2$ for any $\gamma \in \numd$
then the distance is not defined.

Note first that although $\gamma$ is not unique, $\delta(I_1,I_2)$
is well defined: if $I_1=\gamma'I_2$ as well as $I_1=\gamma I_2$
then by proposition \ref{num13} we must have $\gamma'=\epsilon
\gamma=\epsilon_0^k\gamma$ for some $k\in \ZZ$. Hence $\ln
\gamma'=\ln \gamma +kR$.
 Also from eq. (\ref{dist}) we have
$\delta(I_1,I_2)=-\delta(I_2,I_1)$ (when either is defined).

Of particular interest will be the distance $\delta (\co,I)$ of
any principal ideal from the unit ideal $\co$. We write
$\delta(I)$ for $\delta(\co,I)$.

Now recall the principal cycle of reduced ideals
$J_i=\co/\alpha_i=\ZZ+\gamma_i\ZZ$ with $i=0,1,\ldots ,k_0-1$.
Thus $\delta(J_i)=\ln \alpha_i$ and $\delta (J_i,J_k)=\ln\alpha_k
-\ln \alpha_i$. We also had $J_{i+1}=\rho (J_i)$ and
$\alpha_{i+1}=\gamma_i\alpha_i$. Then propositions \ref{num110},
\ref{num111} and \ref{num112} immediately give the following.

\begin{proposition} \label{num33} For all $i\in \ZZ$ \[
\frac{3}{32D} \leq \delta(J_i,\rho(J_i))=\ln \gamma_i\leq
\frac{1}{2}\ln D.
\]
\end{proposition}

\begin{proposition}\label{num34} For all $i\in\ZZ$ \[
\delta(J_i,\rho^2(J_i)) \geq \ln 2. \] \end{proposition}

Furthermore for principal ideals that are not reduced we show that
the reduction process of proposition \ref{num105} leads to a
reduced ideal $I_{\rm red}$ that is close in distance to $I$.

\begin{proposition} \label{num35} Let $I = \redid$ be any (not
necessarily reduced) principal fractional ideal. Let us place
ideals along $\RR$ at positions corresponding to their distances
from $\co$ at 0. In the notation of proposition \ref{num105} let
$i_{\rm red}$ be the least integer such that $I_{\rm
red}=\rho^{i_{\rm red}} (I) = \frac{1}{\alpha}I$ is reduced. Let
$J_k$ be the nearest member of the principal cycle to $I$ with the
property that $\overline{\alpha}_k$ has opposite sign to
$\overline{\alpha}$ (recalling that the $\overline{\alpha}_j$'s
alternate in sign). \\ Then $I$ lies between $J_{k-1}$ and
$J_{k+1}$ and $I_{\rm red}$ is one of $J_{k-1},J_k,J_{k+1}$. Thus
also \[ |\delta (I,I_{\rm red})|<\ln D \] and $\delta(\rho^2
(I_{\rm red}) >\delta(I)$. \end{proposition}

\noindent {\bf Proof}\,\, Recall that $\alpha$ is a minimum of
$I$. The fact that $I$ lies between $J_{k-1}$ and $J_{k+1}$ with
$J_k$ as above, and that $I_{\rm red}$ is one of
$J_{k-1},J_k,J_{k+1}$, follows from the geometrical picture of
ideals as lattices in $\RR^2$ (as claimed in \cite{len82}). \\
------\\ I can't yet quite see how this works and proofs from readers
-- to richard@cs.bris.ac.uk -- would be most welcome! \\
------ \\ Given these facts proposition \ref{num33} gives
$|\delta(I,I_{\rm red})|<\ln D$ and since $\rho$ cycles
consecutively through the principal cycle we get $\delta(\rho^2
(I_{\rm red}) >\delta(I)$. $\qed$

\subsection{Products of ideals -- making large distance
jumps}\label{distance1}

We will need an efficient method of locating ideals (and their
distances) that are far out along the exponentially long principal
cycle. Of course applying $\rho$ repeatedly to $\co$ will
eventually reach any ideal on the cycle (and we can accumulate the
successive distance increments too) but in view of proposition
\ref{num33}, for exponentially distant ideals this will require
exponentially many steps. Hence we introduce a method of
multiplying ideals together (with corresponding addition of
distances) which will allow large distance jumps via iterated
squaring.

If $I_1$ and $I_2$ are ideals then the product $I_1\cdot I_2$ is
defined to be the $\ZZ$-linear span of the set $\{ \alpha\beta:
\alpha\in I_1, \beta\in I_2\}$. $I_1\cdot I_2$ is clearly again an
ideal. If $\{ \alpha_1,\alpha_2\}$ and $\{ \beta_1, \beta_2\}$ are
integral bases of $I_1$ and $I_2$ respectively then $I_1\cdot I_2$
is the $\ZZ$-linear span of $\{ \alpha_1\beta_1,\alpha_1\beta_2,
\alpha_2\beta_1,\alpha_2\beta_2\}$. For principal ideals given as
$I_1=\xi_1 \co$ and $I_2=\xi_2\co$ we simply have $I_1\cdot I_2 =
\xi_1\xi_2 \co$ but we will be interested in computing products of
(reduced) ideals in the presentation of eq. (\ref{frid}).

\begin{proposition} \label{num29}Let \[ I_i=a_i \ZZ +
\frac{b_i+\sqrt{D}}{2}\ZZ \hspace{1cm} i=1,2 \] be principal
ideals. Let $k=\gcd(a_1,a_2,(b_1+b_2)/2)$ and let $u,v,w$ be
integers such that \[ ua_1+va_2+w(b_1+b_2)/2=k \] (which are
guaranteed to exist by Euclid's algorithm). Then \[ I_3=I_1\cdot
I_2 = k\left( a_3\ZZ+\frac{b_3+\sqrt{D}}{2}\ZZ \right) \] where \[
a_3=a_1a_2/k^2 \hspace{1cm} \mbox{and}\hspace{1cm}
b_3=\tau((ua_1b_2+va_2b_1+w\frac{(b_1b_2+D)}{2})/k,a_3).\]
\end{proposition}

\noindent {\bf Proof}\,\, Let
$k(a_3\ZZ+\frac{b_3+\sqrt{D}}{2}\ZZ)$ be the standard
representation of $I_3$ (as in proposition \ref{num18}). For
principal ideals $I=\xi\co$ we had $\cn(I)=|\xi\overline{\xi}|$
(cf.  proposition \ref{num16}) so $\cn(I_3)=\cn(I_1)\cn(I_2)$.
Then using the formula for $\cn(I)$ in proposition \ref{num18} we
get $k^2a_3=a_1a_2$ so $a_3=a_1a_2/k^2$. To derive the claimed
formula for $k$ note first that (cf. proposition \ref{num18})
$b_1^2-D$ is a multiple of 4 so $b_1^2\equiv D \mod 2$. Hence
$b_1\equiv D \mod 2$ (as integers and their squares always have
the same even/odd parity). Similarly $b_2\equiv D\mod 2$ so
$b_1\equiv b_2 \mod 2$ and $(b_1+b_2)/2$ is an integer. Next
recall that $k/2$ is uniquely determined as the least positive
coefficient of $\sqrt{D}$ in any member of $I_3$. But $I_3$ is
generated over $\ZZ$ by $a_1a_2,
a_1(b_2+\sqrt{D})/2,a_2(b_1+\sqrt{D})/2$ and
$(b_1b_2+(b_1+b_2)\sqrt{D}+D)/4$ so the allowable coefficients of
$\sqrt{D}$ have the form $\frac{1}{2}(xa_1+ya_2+z(b_1+b_2)/2)$
with $x,y,z\in\ZZ$. Hence Euclid's algorithm gives
$k=\gcd(a_1,a_2,(b_1+b_2)/2)$. Write \[ k=ua_1+va_2+w(b_1+b_2)/2
\hspace{1cm} \mbox{for $u,v,w\in\ZZ$.}\] Then we must have \[
k\left(\frac{b_3+\sqrt{D}}{2}\right)= ua_1\frac{b_2+\sqrt{D}}{2}
+va_2\frac{b_1+\sqrt{D}}{2}+w \frac{b_1b_2+(b_1+b_2)\sqrt{D}+D}{4}
+sa_1a_2 \] for some $s\in\ZZ$. Equating coefficients of the
$\sqrt{D}$-free terms we get
\[ b_3=\frac{1}{k} \left( ua_1b_2+va_2b_1+w(b_1b_2+D)/2\right)
+2sa_1a_2/k. \] Since $a_1a_2/k=ka_3$ the final term is an even
multiple of $a_3$ and the uniqueness condition $b_3=\tau(b_3,a_3)$
gives the claimed formula. $\qed$

It follows immediately from the definitions that \[
\delta(I_1\cdot I_2)=\delta(I_1)+\delta(I_2) \] if we do not
reduce the sum of the distances $\mod R$. We also point out that
even if $I_1$ and $I_2$ are reduced then $I_1\cdot I_2$ is
generally not reduced.

We will use products primarily in the special case that
$I_1=I_2=I$ where \[ I=\redid = \frac{1}{a}\left(
a\ZZ+\frac{b+\sqrt{D}}{2} \right) \] is a reduced ideal in the
principal cycle. In that case proposition \ref{num29} gives \[ I_2
= I\cdot I= \frac{k'}{a'} \left( a'\ZZ+\frac{b'+\sqrt{D}}{2}\ZZ
\right) \] with \[ k'=\gcd(a,b)=ua+wb \hspace{1cm} a'=a^2/k'^2
\hspace{1cm} b'=\tau((ua+w(b^2+D)/2)/k',a') \] so \[
I_2=\frac{1}{k'}\left(\ZZ + \frac{b'+\sqrt{D}}{2a'}\right)
\hspace{1cm} \mbox{and $\delta(I_2)=2\delta (I)$.} \] Here
$a',b',k'$ are unique and explicitly calculated so we can consider
the ideal \[ I_2'=\ZZ+\frac{b'+\sqrt{D}}{2a'}\ZZ \] which is
generally not reduced. Since $I$ was reduced we have
$0<a,b<\sqrt{D}$ and so $k'=\gcd(a,b)<\sqrt{D}$. Hence
$|\delta(I_2,I_2')|=\ln k'<\frac{1}{2}\ln D$. Next use proposition
\ref{num105} to obtain a reduced ideal $I_2''$ from $I_2'$, which
will be a member of the principal cycle (by theorem \ref{num114}
(c)). Furthermore by proposition \ref{num35} we have
$|\delta(I_2',I_2'')|<\ln D$ so $|\delta
(I_2,I_2'')|<\frac{3}{2}\ln D$. Hence in view of proposition
\ref{num34} if we apply $\rho$ or $\rho^{-1}$ to $I_2''$, \, $2n$
times where $n<(\frac{3}{2}\ln D)/\ln 2 = O(\ln D)$, then we can
locate the first member $J_k$ of the principal cycle with distance
$\delta(J_k)>2\delta(I)$ (and also compute its distance
$\delta(J_k)$). We denote this uniquely determined ideal $J_k$ by
$I*I$ i.e. for any member $I$ of the principal cycle, $I*I$ is the
first member with distance exceeding twice the distance of $I$.
(In the above construction if $2\delta(I)$ exceeds $R$ then we
simply wrap around the principal cycle, passing through $\co$
again.)

Let us now estimate the computational complexity of computing
$I*I$ from $I$. Since $I$ was reduced we have $a,b=O(\sqrt{D})$.
Also in $k'=ua+wb$ the integers $u,w$ are not unique and we have
the freedom (leaving $k'$ unchanged): \[ u\rightarrow u-xb
\hspace{1cm} w\rightarrow w+xa \hspace{1cm} \mbox{for any
$x\in\ZZ$.}\] Hence we can always make $u<b=O(\sqrt{D})$ and
$k'=ua+wb$ gives $w=O(D)$ too. Thus $a',b',k'$ are all at most
$O(D^2)$ and the arithmetic calculation of $I_2'$ can be performed
in $O(\poly \log D)$ time. Furthermore the reduction operation of
proposition \ref{num105} requires $O(\log(a'/\sqrt{D}))$ steps for
$a'=O(D^2)$ and to compute $I_2''=I*I$ we apply $\rho^{-2}$ or
$\rho^2$ at most $(\frac{3}{2}\ln D)/\ln 2 = O(\ln D)$ times.
Hence the entire computation of $I*I$ from $I$ can be performed in
$O(\poly \log D)$ time and we have proved the following.

\begin{proposition} \label{num120} Let $I$ be a reduced principal ideal.
Consider the iterated $*-$squaring: \[ I\rightarrow I^{(2)}=I*I
\rightarrow I^{(4)}=I^{(2)}*I^{(2)} \rightarrow \cdots \rightarrow
I^{(2^n)}=I^{(2^{n-1})}*I^{(2^{n-1})}.\] The final ideal
$I^{(2^n)}$ has distance $\delta(I^{(2^n)})>2^n \delta(I)$ (where
we have not reduced the growing distances $\mod R$) and this ideal
(with its unreduced distance) can be computed in $O(\poly(\log
D),n)$ time.
\end{proposition}

More generally if $I_1$ and $I_2$ are reduced ideals \[
I_i=\ZZ+\frac{b_i+\sqrt{D}}{2a_i}\ZZ \] (so $0<a_i,b_i <\sqrt{D}$)
we define $I_1*I_2$ to be the first member of the principal cycle
whose distance exceeds $\delta(I_1)+\delta(I_2)$. Then following
the methods above it is easy to see that $I_1*I_2$ (i.e. its $a,b$
parameters) can be computed in $\poly \log D$ time from
$a_1,a_2,b_1,b_2$.

\section{Summary -- the picture so far} \label{summary}

Given $d$ we have the algebraic integers $\co \subset \numd$ and a
finite (but exponentially large in $\log d$) set of principal
reduced ideals $\{ J_0=\co, J_1, \ldots, J_{k_0-1}\}$ of $\co$.
Each $J_i$ is defined by a pair of integers $a,b$ with $0\leq a,b
<\sqrt{D}$ and hence has a $\poly \log d$ sized description. We
write $J_i=\ZZ+\gamma_i\ZZ =\ZZ +\frac{b_i+\sqrt{D}}{2a_i}\ZZ$.

The reduced ideals $J_i$ can be placed around a circle of
circumference $R=\ln \epsilon_0$ at irregularly spaced distances
$\delta(J_i)$ from the ideal $\co$ at distance 0, giving the
so-called principal cycle. Since $J_i$ is principal it has the
form $J_i=\alpha_i \co$ for some $\alpha_i \in \numd$ and then
$\delta (J_i)=\ln |\alpha_i| \mod R$.

We have a reduction operator $\rho$ such that $J_{i+1}=\rho(J_i)$
and $\rho$ and $\rho^{-1}$ are polynomial time computable mappings
allowing us to step either direction around the principal cycle.

We have an (exponentially small) lower bound $d_{\rm min}=3/(32D)$
on the distance between consecutive ideals $J_i$ and an upper
bound of $\frac{1}{2}\ln D$. We also have a constant lower bound
of $\ln 2$ on the distance between $J_i$ and $J_{i+2}$.

If $J_i$ is given in the form $\ZZ+\gamma_i\ZZ$ then we cannot
compute $\delta (J_i)$ in $\poly \log d$ time (e.g. we cannot
translate $\ZZ+\gamma_i\ZZ$ efficiently to the $\alpha_i\co$
description). But we can efficiently compute the distance
increments induced by $\rho$ and obtain the distance increment
from $J_i$ to $J_{i+1}$ in time $\poly \log d $ if $\gamma_i$ is
given.

The above method does not allow us to efficiently move far out
along the principal cycle starting from $\co$ (as this would
require exponentially many steps). To achieve such large jumps we
introduced a product operation $I*J$ on reduced ideals $I,J$ with
the property that $I*J$ is the first reduced ideal having distance
exceeding $\delta(I)+\delta(J)$. $I*J$ and $\delta(I*J)$ are then
efficiently computable so by iterated $*-$squaring of $J_2=\rho^2
(\co)$ having $\delta(J_2)>\ln 2$ we can move out to a distance
exceeding $2^n\ln 2$ in $\poly (\log d, n)$ time. If $2^n\ln 2$
exceeds $R$ as $n$ increases, the increasingly distant ideals
simply wrap around the circle while the distance we compute is not
reduced $\mod R$.

\section{The periodic function for Hallgren's algorithm}
\label{hall-per} Recall that $\cp\ci_{\rm red}$ denotes the
(finite) set of all reduced principal fractional ideals i.e. the
principal cycle,  and $\delta(I)$ denotes the distance of $I$ from
the unit ideal $\co$. Define $h:\RR \rightarrow \cp\ci_{\rm red}
\times \RR$ as follows. Let $\tilde{x}\equiv x \mod R$ with $0\leq
\tilde{x}<R$. Then \[ h(x)=(I_x, \tilde{x}-\delta(I_x)) \] where
$I_x \in \cp\ci_{\rm red}$ is the ideal having greatest distance
$\delta(I_x)<\tilde{x}$. In other words, if we place ideals $I$
along the real axis at positions in $[0,R)$ given by their
distances $\delta(I)$ from $\co$ at $x=0$, and periodically
reproduce this pattern in each interval of length $R$, then $I_x$
is the ideal that is nearest to the left of $x$ and
$\tilde{x}-\delta(I_x)$ is the distance gap up to $x$.

\noindent {\bf Remark} Intuitively we simply wanted $h'(x)=I_x$
but we include the information of the positive distance gap
$\tilde{x}-\delta(I_x)$ in the value of $h(x)$ to ensure that $h$
is a one-to-one function within each period (noting that $h'$ is
constant for $x$ varying between consecutive ideals).

The main point of our whole development of the theory of reduced
ideals is to prove the following result.

\begin{theorem}\label{num36} The function $h$ is computable in
polynomial time. More precisely, if  $x$ is an integer multiple of
$10^{-n}$ we can compute the ideal $I_x$ exactly and an
approximation of $\tilde{x}-\delta(I_x)$ accurate to $10^{-n}$ in
time poly$(\log D, \log x, n)$.\\
Also $h$ is a periodic function with period $R$ and it is
one-to-one within each period.\end{theorem}

\noindent {\bf Proof}\,\,  Clearly from the definition, $h$ is
periodic and one-to-one within each period.

Given $x$ we can compute the ideal $I_x \sim (a,b)$ as follows.
Recall that the action of $\rho$ and $\rho^{-1}$ on a reduced
ideal can be computed in $\poly \log D$ time. We start with $\co
\sim (\tau(D,2),1)$ at $\delta(\co)=0$. Apply $\rho$ twice to get
$I_0=\rho^2 (\co)$ with $\ln D > \delta (I_0)>\ln 2$. Write
$A=\delta (I_0)$. Compute a sequence of ideals $I_0, I_1, I_2,
\ldots $ by repeated $*-$squaring i.e. $I_{k+1}=I_k*I_k$. Hence \[
2\delta(I_{k-1})\leq \delta (I_k) \leq
2\delta(I_{k-1})+\frac{1}{2}\ln D \] and so $\delta (I_k)\geq
A2^k$ (and we are not reducing the distance $\mod R$). We
terminate the iteration with $k=N$ when $\delta (I_{N+1})$ first
exceeds $x$, which always happens for $N<\lceil \log_2
(x/A)\rceil$.

Now take $I_N$ and continue moving towards $x$ from the left in
successively smaller steps, first using $*-$products with
$I_{N-1}$, then $I_{N-2}$ etc, ensuring each time to stay to the
left of $x$. More precisely if we are at a reduced ideal $J$ and
we are using $I_k$ then we compute $J*I_k$ with
$\delta(J)+\delta(I_k)\leq \delta (J*I_k)\leq
\delta(J)+\delta(I_k)+\frac{1}{2}\ln D $ and replace $J$ by
$J*I_k$ so long as $\delta(J*I_k)<x$. If $\delta(J*I_k)>x$ we
discard it and try $J*I_{k-1}$. Finally after $O(N)=O(\log x)$
steps, finishing with $I_1$ we will have computed a reduced ideal
$J_*$ to the left of $x$ with $x-\delta(J_*)\leq A+\frac{1}{2}\ln
D \leq \frac{3}{2}\ln D$.

While doing all this with the ideals we also carry along a
parallel computation of their accumulating distances (but not mod
$R$) using the formula $\delta (I,\rho(I))=\ln |\gamma|$ for
$I=\ZZ+\gamma\ZZ$ and proposition \ref{num120}, calculating to a
sufficient accuracy that will give the final $\delta(J)$ (at the
end of our process continuing below) to the desired accuracy.

Finally we repeatedly apply $\rho^2$ to $J_*$ until the distance
again exceeds $x$. This will require at most $(\frac{3}{2}\ln
D)/\ln 2$ steps. If $J'$ is the last such ideal to the left of $x$
then $I_x$ will certainly be either $J'$ or $\rho(J')$ which we
can determine by finally checking if $\delta(\rho(J'))$ exceeds
$x$ or not.

As mentioned above, while doing all this with the ideals we also
carry along a parallel computation of their accumulating
distances. Since $n$ digit arithmetical operations are computable
with poly$(n)$ effort, this whole process gives the full value of
$h(x)$ -- with $I_x$ precisely and $\tilde{x}- \delta(I_x)$ to $n$
digits of accuracy -- with poly$(\log D,\log x, n)$ effort. By
proposition \ref{num33}, if  $10^{-n} <d_{\rm min}=3/(32D)$ (a
lower bound on the minimum distance between ideals), we will
certainly have located the {\em nearest} form to the left of $x$.
(If $n$ is not sufficiently large, we still always work to an
accuracy which is at least as fine as the above bound.) $\qed$

\noindent {\bf Proof of proposition \ref{num3}}\,\, Given $\lceil
R\rceil$ or $\lfloor R\rfloor$ we use theorem \ref{num36} to
compute the closest ideal $I$ to the left of $\lfloor R\rfloor$
and also its distance (to any desired accuracy). Thus
$\delta(\rho(I))
>\lfloor R\rfloor$ and $\delta(\rho^3(I))>\lfloor R \rfloor +\ln 2
>\lceil R\rceil$. It follows that either $\rho(I)$ or $\rho^2(I)$
must be $\co = J_0$ and its distance (that we compute) gives $R$.
$\qed$

\section{The quantum algorithm for irrational periods on $\RR$}
\label{quantum}

Suppose we have a function on $\RR$ that is periodic with
(possibly irrational) period $R$
\[ f:\RR \rightarrow X \hspace{1cm} f(x+R)=f(x) \hspace{5mm}
\mbox{for all $x\in \RR$.} \] To apply the quantum period finding
algorithm we will need to suitably discretise $f$, by taking
values that are integer multiples $k/N$ of $1/N$ (for suitably
large $N$) and if $X$ also contains continuous variables it should
be discretised too, to ensure that exact calculations can be
performed.

For example if $f:\RR\rightarrow \RR$ then we could define
\[ \tilde{f}:\ZZ \rightarrow \frac{1}{N}\ZZ \] by
\[ \tilde{f}(k)=\lfloor f(\frac{k}{N})\rfloor_N \]
where we use the notation $\lfloor x\rfloor_N$ to denote the value
of $x$ rounded down to the nearest multiple of $1/N$ (and
similarly $\lceil x \rceil_N$ for rounding up). We would want
$\tilde{f}$ to contain suitable approximate information about the
period $R$ but unfortunately this is not guaranteed: suppose that
$f$ has a very large variation in the region of diameter $1/N$
around $x=k/N$. Then although $f(k/N)=f(k/N+lR)$ exactly for all
$l\in\ZZ$, if we round $lR$ down (or up) to the nearest multiple
of $1/N$ then the values of $f(k/N+\lfloor lR\rfloor_N)$ could
vary arbitrarily with $l$ because (for irrational $R$) the
rounding gap $0\leq lR-\lfloor lR\rfloor_N \leq 1/N$ is generally
dense in the interval $[0,1/N]$ as $l$ ranges over $\ZZ$. Thus the
periodicity may not be evident (even approximately) in
$\tilde{f}$. To rule out such behaviour, we introduce the
following notion of ``weak periodicity'' which will suffice for
our applications.

\noindent {\bf Definition} A function $f:\ZZ \rightarrow X$ is
called {\bf weakly periodic} with period $S\in \RR$ if for each
$0\leq k\leq \lfloor S\rfloor$ and each $l\in\ZZ$ \beq
\label{wper} \mbox{either $f(k+\lfloor lS \rfloor )$ or
$f(k+\lceil lS \rceil )$ equals $f(k)$}. \eeq We write $f(k)=
f(k+[lS])$ where the notation $[lS]$ denotes a chosen one of the
two values $\lfloor lS\rfloor$ or $\lceil lS\rceil$ (and the
choice may vary with $k$ and $l$).

\noindent {\bf Remark} In our applications $f$ with period $S$
will arise from a computational problem with some input size
$\sigma$ and $S$ will grow as $\log S=O(\poly (\sigma))$. Then we
will require the condition eq. (\ref{wper}) to hold only for a
suitably large fraction $1-\frac{1}{\poly(\sigma)}$ of the values
$0\leq k\leq \lfloor S\rfloor$. $\qed$

Recall our fundamental function with period $R$ (the regulator):
\[ h:\RR\rightarrow \cp\ci_{\rm red} \times \RR \hspace{1cm}
h(x)=(I_x, \tilde{x}-\delta(I_x)) \] where $I_x$ is the reduced
principal ideal closest in distance to the left of $x$ and $
\tilde{x}-\delta(I_x))$ is the distance gap from $I_x$ to $x$.
Define \[ \tilde{h}_N:\ZZ \rightarrow \cp\ci_{\rm red} \times
\frac{1}{N}\ZZ \hspace{1cm} \tilde{h}_N(k)= (I_{k/N},\lfloor
k/N-\delta(I_{k/N})\rfloor_N) \] i.e. we compute $h(x)$ for
$x=k/N$ and round the distance gap value down to the nearest
integer multiple of $1/N$.

\begin{proposition}\label{num41}  (i) $\tilde{h}_N$ is
one-to-one on $0\leq k\leq \lfloor NR \rfloor$.\\ (ii)
$\tilde{h}_N(k)$ is computable in $\poly (\log k, \log N,\log d)$
time, so if $N$ and $k$ are $O(\poly (d))$ then $\tilde{h}_N(k)$
is computable in time $\poly(\log d)$.\\ (iii) Let $d_{min} =
3/(32D)$ be a lower bound on the minimum distance between reduced
ideals, and recall that $\sigma=\log d$ is the input size for the
computation. If $N$ is sufficiently large i.e. $1/N < d_{min}
/\log d$ then $\tilde{h}_N$ is weakly periodic with period $NR$.
In fact the condition eq. (\ref{wper}) holds at all values $0\leq
k\leq \lfloor NR\rfloor$ except possibly at the largest multiple
$k/N$ of $1/N$ to the left of each reduced ideal (which is at most
a fraction $1/\log d$ of the values).
\end{proposition}

\noindent {\bf Proof}\,\,  (i) $\tilde{h}_N(k)=\tilde{h}_N(l)$
means that $k/N$ and $l/N$ have the same nearest ideal on the left
and the distances to it (from $k/N,l/N$) are the same when rounded
down to the nearest multiple of $1/N$. But $|k/N-l/N|\geq 1/N$ if
$k\neq l$ so the rounded distances must then be different. Hence
$\tilde{h}_N(k)=\tilde{h}_N(l) \Rightarrow k=l$.\\ (ii) This is an
immediate consequence of theorem \ref{num36} noting also that
arithmetic operations with integers of size $O(N)$ (such as
rounding operations) can be performed in $\poly (\log N)$ time.\\
(iii) Consider any fixed value $0\leq k\leq \lfloor NR\rfloor$.
Let $I'$ at distance $\delta(I')=x_0\leq k/N$ be the nearest ideal
to the left of $k/N$, with distance gap $n_0/N+\epsilon_1$ where
$n_0\in \NN$ and $0<\epsilon_1<1/N$. Since $h$ is exactly
periodic, $I'$ is also the nearest ideal to the left of $k/N+lR$,
having the same distance gap. But now $k/N+lR$ is not an exact
multiple of $1/N$ so consider rounding it up and down to $\lfloor
k/N+lR\rfloor_N$ and $\lceil k/N+lR\rceil_N$. Let the
corresponding rounding distances be $\epsilon_2$ (down) and
$\epsilon_3$ (up). Thus $\epsilon_2+\epsilon_3=1/N$. We then have
the following unrounded distances back to $I'$:\\ from
$k/N$:\hspace{3mm} $n_0/N+\epsilon_1$; \\ from $\lfloor
k/N+lR\rfloor_N$:\hspace{3mm}
$n_0/N+\epsilon_1-\epsilon_2= n_0/N+\epsilon_1 -1/N+\epsilon_3$;\\
from $\lceil
k/N+lR\rceil_N$:\hspace{3mm} $n_0/N+\epsilon_1+\epsilon_3$.\\
Hence if $\epsilon_1+\epsilon_3>1/N$ then $k/N$ and $\lfloor
k/N+lR\rfloor_N$ will have the same rounded distances and
$\tilde{h}_N(k)=\tilde{h}_N(k+\lfloor lR\rfloor)$. (Note that the
rounded down position $\lfloor k/N+lR\rfloor_N$ can never pass to
the left of $I'$ because $\epsilon_1+\epsilon_3>1/N$ so the
unrounded distance $n_0/N+\epsilon_1-1/N+\epsilon_3>n_0/N$).\\ If
$\epsilon_1+\epsilon_3\leq 1/N$ then $k/N$ and $\lceil
k/N+lR\rceil_N$ will have the same rounded distances to $I'$. But
rounding $k/N+lR$ {\em up} to $\lceil k/N+lR\rceil_N$ may cross
over another ideal $I''$ located in the gap between these two
values. This can happen only if $k/N$ is the greatest multiple of
$1/N$ to the left of $I''$. Thus, except for this eventuality, we
have $\tilde{h}_N(k)=\tilde{h}_N(k+\lceil lR\rceil)$. \\ In
summary, the weak periodicity condition eq. (\ref{wper}) must hold
at all $k$ except possibly those $k$ with $k/N$ being the largest
such value to the left of a reduced ideal. Now if $1/N<d_{min}
/\log d$ i.e. $N>\log d /d_{min}$ then we have at least $\log d$
points between any two ideals so the weak periodicity condition
will hold at least for a fraction $(\log d-1)/\log d = 1-1/\log d$
i.e. $1-1/\poly (\log d)$, of all $k$ values. $\qed$

\begin{theorem}\label{num42} Suppose that $f:\ZZ\rightarrow X$ is
weakly periodic with period $S$ and \\ (a) $f(k)$ is computable in
$\poly (\log k,\log S)$ time;\\ (b) $f$ is one-to-one for $0\leq
k\leq \lfloor S\rfloor$; \\ (c) given an integer $m$ there is a
$\poly(\log S)$ time algorithm that will test if $m$ is close to
an integer multiple of $S$ or not i.e. if $|jS-m|<1$ for some
$j\in \ZZ$ or not.\\ Then there is a quantum algorithm with
running time $\poly(\log S)$ that outputs an integer $a$ with
$|S-a|<1$ with probability $\geq 1/\poly (\log S)$.\end{theorem}

\noindent {\bf Proof}\,\,  Introduce some further notation:
$\lfloor x\rceil$ for the nearest integer above or below $x$. Thus
$|x-\lfloor x\rceil| \leq 1/2$. \\ We will use quantum Fourier
sampling in a dimension $q$ with $q\geq 3S^2$ (cf. later for the
origin of this choice) and $q$ a power of 2 (for ease of efficient
implementation of the quantum Fourier transform). Construct the
state \[ \frac{1}{\sqrt{q}} \sum_{m=0}^{q-1} \ket{m}\ket{f(m)} \]
which by (a) can be done in $\poly(\log S)$ time. Write \[ q=pS+r
\hspace{1cm} p,r\in \ZZ \hspace{5mm} 0\leq r< S \] i.e. $pS$ is
the largest multiple of $S$ that is $\leq q$ so $pS\leq q$.
Measure the second register to obtain in the first register:
\[ \ket{\psi_0}= \frac{1}{\sqrt{p}} \sum_{i=0}^p \ket{k+[iS]} \]
for $0\leq k\leq \lfloor S\rfloor$ chosen uniformly. \\ (Note:
here we have assumed that the weak periodicity condition eq.
(\ref{wper}) holds for all $k$. If it holds only for a fraction
$1-1/\poly (\log S)$ of the $k$ values, then the state
$\ket{\psi_0}$ will be slightly modified and our estimates below
will be altered by a suitably small ($1/\poly (\log S)$) amount.
Our final conclusions will remain valid but for clarity we will
omit explicit analysis of these extra obfuscating variations.)\\
Next apply the quantum Fourier transform $\mod q$ to
$\ket{\psi_0}$ to obtain the state $\sum_j a_j \ket{j}$. We will
be interested in the output probabilities $|a_j|^2$ and these do
not depend on $k$ (by the shift invariance property of the Fourier
transform). Hence (wlog) we will set $k=0$. Then \[
a_j=\frac{1}{\sqrt{pq}}\sum_{l=0}^{p-1} e^{2\pi i \frac{j[lS]}{q}}
.
\] Write $[lS]=lS+\delta_l$ with $-1<\delta_l<1$. Consider those
$j$'s that are nearest to an integer multiple of $q/S$: \beq
\label{j1} j=\lfloor \frac{kq}{S}\rceil = \frac{kq}{S}+\epsilon
\hspace{5mm} \mbox{$0\leq k\leq S $ and $-\frac{1}{2}\leq \epsilon
\leq \frac{1}{2}$} \eeq and also those $j$'s that are not too
large: \beq \label{j2} j<\frac{q}{\log S}. \eeq For these $j$'s we
have
\[ \frac{j[lS]}{q}=
(\frac{k}{S}+\frac{\epsilon}{q})(lS+\delta_l)= \frac{\epsilon
S}{q}+( \frac{k\delta_l}{S}+\frac{\epsilon\delta_l}{q}) \mod 1 \]
where we have removed the integer $\frac{k}{S}lS$. Since we are
taking $j=kq/S+\epsilon <q/\log S$ and $|\epsilon | \leq 1/2$ we
get $k/S<1/\log S+1/(2q)$. Also $|\delta_l|<1$ and $q>3S^2$ so \[
|\frac{k\delta_l}{S}+\frac{\epsilon \delta_l}{q}| < \frac{1}{\log
S}+\frac{1}{2q}+\frac{1}{2q}\leq \frac{2}{\log S}. \] Then writing
$A=\epsilon Sp/q$ we have \[ a_j = \frac{1}{\sqrt{pq}}
\sum_{l=0}^{p-1} e^{2\pi i (\frac{Al}{p}+\xi(l))} \] where \[
\xi(l)=\frac{k\delta_l}{S}+\frac{\epsilon\delta_l}{q}\hspace{5mm}
\mbox{has $|\xi(l)|\leq \frac{2}{\log S}$}\] and $A=\epsilon Sp/q$
has (recalling $pS<q$) \[ |A| =|\frac{\epsilon Sp}{q}|\leq
|\epsilon| \leq \frac{1}{2}. \] Hence by lemma \ref{num44} below
there is a constant $c$ such that \[ |a_j|^2 \geq \frac{1}{pq}cp^2
= \frac{cp}{q}\geq \frac{cp}{pS} = \frac{c}{S} \] i.e. for each
$j$ satisfying eqs. (\ref{j1}) and (\ref{j2}) we have \[
prob(j)\geq \frac{c}{S} \] and they are uniformly distributed. How
many such $j$'s are there? We have $ j=\lfloor \frac{kq}{S}\rceil
\leq q/ \log S$ so $0\leq k\leq S/\log S$. Thus the probability of
getting a $j$ value that satisfies eqs. (\ref{j1}) and (\ref{j2})
is $\geq c/\log S$.

Running all the above {\em twice} we will obtain two such $j$
values (called $c$ and $d$): \beq \label{klcd} c=\lfloor
\frac{kq}{S}\rceil \hspace{1cm}d=\lfloor \frac{lq}{S}\rceil \eeq
having $\gcd (k,l)=1$ with probability $1/\poly(\log S)$. (The
$\gcd (k,l)=1$ condition is obtained with inverse polynomial
probability by the prime number theorem). From $c$ and $d$ we want
to extract the information of $k$. To do this we use properties of
continued fractions: we show that $k/l$ is a convergent of the
continued fraction of $c/d$, which then gives $k$ as a numerator
of a convergent. We use the following basic property of continued
fractions (cf. theorem 184 of \cite{hw}): If $a,b\in \NN$ and
$|x-\frac{a}{b}|\leq \frac{1}{2b^2}$ then $a/b$ is a convergent of
the continued fraction of $x$. By lemma \ref{num43} below (which
requires $q>3S^2$) we have \[ \left|\frac{c}{d}-\frac{k}{l}\right|
< \frac{1}{2l^2} \] giving the required result.

Recall that $c=\lfloor kq/S\rceil$. Then for each convergent
$c_n/d_n$ of the continued fraction of $c/d$ we check if $c_n=k$
by computing $m=\lfloor c_n q/c\rceil$ and using (c) (from the
statement of the theorem) to check if it is within 1 of an integer
multiple of $S$ or not. Then we output the smallest such $m$ that
is within 1 of an integer multiple of $S$ (and with probability
$1/\poly(\log S)$ this multiple is 1). The various rounding
processes stay within the required accuracy because if $c=\lfloor
kq/S \rceil$ and $q\geq 3S^2$ then $|S-\lfloor kq/c\rceil| \leq
1$. (To see this write $c=kq/S+\epsilon$ with $|\epsilon| \leq
1/2$. Then
\[ \frac{kq}{c}=\frac{S}{1+\frac{\epsilon S}{kq}}=
\frac{S}{1+\alpha}
\] where $|\alpha|<1/(6S)$ for $q\geq 3S^2$. Thus $kq/c=S-S\alpha
/(1+\alpha)$ and $|S\alpha/(1+\alpha)|<1/2$.)

The result of the whole process above is to output an integer $m$
such that $|S-m|<1$ with probability $1/\poly(\log S)$. $\qed$

\begin{lemma}\label{num43} If $q>3S^2$ then \[
\left|\frac{c}{d}-\frac{k}{l}\right| <\frac{1}{2l^2} \] with
$k,l,c,d$ as in eq. (\ref{klcd}).\end{lemma}

\noindent {\bf Proof}\,\,  Let $c=kq/S+\epsilon_k$ and
$d=lq/S+\epsilon_l$ with $|\epsilon_k|$ and $|\epsilon_l|$ both
$\leq 1/2$ and wlog take $k<l\leq S$. Then \[
\left|\frac{c}{d}-\frac{k}{l}\right| =
\left|\frac{S(\epsilon_kl-\epsilon_l
k)}{l^2q+\epsilon_lSl}\right|\leq
\left|\frac{S(l+k)}{2l^2q-2Sl/2}\right|\leq \frac{S}{lq-S/2}\]
where in the second last inequality we have used the worst case
$\epsilon_k=1/2$ and $\epsilon_l=-1/2$. Finally note that \[
\frac{S}{lq-S/2}\leq \frac{1}{2l^2} \] holds if $q\geq 2lS+S/(2l)$
so $q\geq 3S^2$ suffices (as $l\leq S$). $\qed$

\begin{lemma} \label{num44} Let $|A|\leq 1/2$ and let $\xi(l)$ be
any function satisfying $|\xi(l)|<1/n$ with $n \sim O(\log p)$.
Then there is a constant $c$ such that for all sufficiently large
$p$: \[ X=\left|\sum_{l=0}^{p-1} e^{2\pi i
(\frac{A}{p}l+\xi(l))}\right|^2 \geq cp^2.\] \end{lemma}

\noindent {\bf Proof}\,\,  View $b_l=\exp 2\pi i (Al/p+\xi(l))$ as
points on the unit circle, being $\xi(l)$-perturbations of the
evenly spaced points $c_l=\exp 2\pi i Al/p$ for $l=0, \ldots ,p-1$
which range over a fraction $|A|\leq 1/2$ of the whole circle.

Introduce $x,y$ axes so that the $c_l$ points are mirror symmetric
in the $y$ axis with the negative $y$ axis bisecting the unused
part of the circle. For all sufficiently large $p$ it is clear
that the total $y$ component of $\sum b_l$, for {\em any}
perturbation with $\xi(l)<1/n$, is positive (since most points
will lie in the half circle having $y\geq 0$). We claim that the
smallest value of $X$ (over all possible perturbations) will occur
when $\xi(l)=1/n$ for all points $c_l$ having $x<0$ and
$\xi(l)=-1/n$ for all points having $x>0$ i.e. $\xi(l)$ rotates
points away from the positive $y$ axis, down towards the negative
$y$ axis, symmetrically on the two sides of the $y$ axis. This
perturbation maintains zero total $x$ component (by symmetry) and
hence has the least squared total $x$ component of all
perturbations, and for each point we get the least possible
positive, or most negative $y$ component amongst all
perturbations. Hence the total $y$ component (always being
positive) must attain its least possible value and so we have the
least possible $X$ amongst all perturbations.

To obtain a lower bound on $X$ for this minimal perturbation note
that since $|A|\leq 1/2$, negative $y$ values can occur (if at
all) only in the arc of the circle of fraction $1/n$ below the
$\pm x$ axes, which are counterbalanced by positive values in the
corresponding arcs above the $\pm x$ axes. Hence all points in the
arc $(2\pi/n,\pi-2\pi/n)$ will contribute uncancelled positive $y$
values so in particular \[ X\geq [\mbox{number of points in arc $
(\pi/4,3\pi/4)$}]^2 \cdot [\mbox{$y$ component at angle
$\pi/4$})]^2.
\] Since $|A|\leq 1/2$ the number of points in $(\pi/4,3\pi/4)$ is
at least $p/3$ (actually it's only slightly less than $p/2$ for
large $n$) and the $y$ component at $\pi/4$ is $1/\sqrt{2}$ so $X
\geq (p^2/9)(1/2) = p^2/18$ for all sufficiently large $p$. $\qed$

Finally we can apply theorem \ref{num42} to $\tilde{h}_N$ of
proposition \ref{num41} having $S=NR$ to get an integer $m$ with
$|NR-m|<1$ i.e. the value of $R$ to a tolerance of $1/N$. The
validity of theorem \ref{num42}'s requirement (c) in this case can
be seen as follows.

Given an integer $m$ , to test if $|jR-m|<1$ or not, for some
$j\in \ZZ$, we first compute $I_m$, the ideal closest to $m$ on
the left (which can be done in $\poly (\log m, \log d)$ time (cf.
theorem \ref{num36}). Let $I_m$ be $J_{i_0}$ in the principal
cycle. Now $|jR-m|<1$ for some $j$ iff the ideal $\co$ (in the
repeating periodic pattern of reduced ideals placed along $\RR$)
is located at distance $<1$ from $m$. Recall also that $\delta
(J_i,J_{i+2})>\ln 2$ and $2\ln 2 >1$. Hence if we look at the
ideals $J_{i_0-4},J_{i_0-3}, \ldots ,J_{i_0+4}$ (which can be
efficiently constructed by applying $\rho$ and $\rho^{-1}$ to
$I_m$, up to four times) we will be able to see if $\co$ lies
within distance 1 of $m$.

Putting all this together we have proved:

\begin{theorem} \label{num45} Given a square-free integer $d\in
\NN$ there is a quantum algorithm that will output the regulator
$R$ of $\numd$ to accuracy $10^{-n}$ with running time $\poly(\log
d,n)$ and success probability $1/\poly(\log d,n)$, so long as
$10^{-n}$ is sufficiently small: $10^{-n}<d_{min}/\log d$.
\end{theorem}

\noindent In view of proposition \ref{num3}, to get $R$ to
accuracy $10^{-n}$ it suffices to use the quantum algorithm in
theorem \ref{num42} with a {\em fixed} (suitably large) value
$n_0$ of $n$ so the success probability to get accuracy $10^{-n}$
becomes $1/\poly (\log d)$ but (as expected) the running time
remains $\poly (\log d,n)$.

\section{Further Remarks}\label{further}

Hallgren's algorithm computes the regulator of $\numd$ in $\poly
\log d$ time. The tenacious reader may wish to estimate the degree
of the polynomial running time by assessing all the ingredients
that we have described in detail.

It is interesting to compare the computational complexity of the
task REG -- computing the regulator of $\numd$ -- to that of the
task FAC -- factoring a given integer $d$ (especially as the
latter also admits an efficient quantum algorithm {\em viz.}
Shor's algorithm). Write $n=\log d$. The best classical algorithms
for FAC and REG are sub-exponential but super-polynomial, with
running times $\exp (O(n^{1/3}))$ and $\exp (O(n^{1/2}))$
respectively. Thus REG is the harder task and indeed there is a
known reduction of FAC to REG (i.e. an algorithm for REG can be
used to achieve FAC with the same time complexity). However it is
significant to note that the sub-exponential algorithms
\cite{will02} (\S 8) for REG depend on the truth of a suitable
generalised Riemann hypothesis (GRH) associated with zeta
functions on the quadratic number field. Without this assumption
the best algorithm for REG has exponential running time
$O(d^{1/4})$ (in fact using the same mathematical formalism of
reduced ideals that we presented). Furthermore FAC is in ${\rm NP}
\bigcap {\rm co}-{\rm NP}$ whereas REG is known to be in NP only
under the assumption that GRH is true.

Finally we mention that in addition to solving Pell's equation,
Hallgren's algorithm may be readily adapted to give efficient
solutions of two further fundamental problems of computational
algebraic number theory (cf. \cite{coh93}): the principal ideal
problem and the computation of the so-called class group of
$\numd$ (and its size, the class number). We refer the reader to
Hallgren's paper \cite{hallgren} for a description of these
further applications and here we only make a few brief remarks
about the statement of these problems.

For the usual integers $\ZZ \subset \QQ$ all ideals are principal
ideals. However in the more general setting of the algebraic
integers $\co$ in the quadratic number field $\numd$ this is no
longer true i.e. there exist ideals of $\co$ which are not of the
form $\alpha\co$. However proposition \ref{num14} remains true:
any ideal has the form $I=\alpha\ZZ+\beta\ZZ$ (for suitable
$\alpha,\beta \in \numd$ which are restricted by the requirement
that $I$ be closed under multiplication by $\co$). The principal
ideal problem is then: given an ideal as $I=\alpha\ZZ+\beta\ZZ$
determine whether it is a principal ideal (and if it is, compute a
generator $\gamma$ such that $I=\gamma\co$).

The class group of $\numd$ provides a measure of how much the
ideals of $\co$ can deviate from being principal. An ideal $I$ is
called invertible if there exists an ideal $J$ such that $I\cdot
J=\co$ (using the product of ideals introduced in section
\ref{distance1}). Clearly all principal ideals are invertible (as
the inverse of $\alpha\co$ is $\frac{1}{\alpha}\co$) and the set
$\ci_{\rm inv}$ of all invertible ideals is an abelian group under
multiplication of ideals (with $\co$ being the identity). The
subset $\cp$ of all principal ideals is a subgroup and the class
group $\cc$ is defined to be the quotient $\cc=\ci_{\rm inv}/\cp$.
Now it can be shown that $\cc$ is always a {\em finite} abelian
group (cf. \cite{coh93}) and the class group problem is to compute
a set of generators of $\cc$ and to compute the size of $\cc$.

Much of the theory of reduced ideals that we developed for
principal ideals can be readily extended to general ideals
providing a tool for attacking these problems too. Furthermore
there is a way of representing ideals in terms of binary quadratic
forms on $\ZZ$. Roughly speaking, the ideal
$\ZZ+\frac{b+\sqrt{D}}{2a}\ZZ$ is represented by the form
$ax^2+bxy+cy^2$ where $D=b^2-4ac$ and $x,y\in\ZZ$, and we can
develop a corresponding theory of reduced forms. (See \cite{coh93}
for an exposition of the correspondence between ideals and
quadratic forms). Gauss devoted much effort to the class group
problem when formulated in these terms, before the introduction of
the concept of ideals by E. Kummer in 1847.

\noindent {\large\bf Acknowledgements}\\ RJ \label{ackn} is
supported by the UK Engineering and Physical Sciences Research
Council. Much of this work was carried out during the Quantum
Computation Semester (Fall 2002) at the Mathematical Sciences
Research Institute, Berkeley California and the author is grateful
for the hospitality of the MSRI. Finally, thanks to Sean Hallgren
and Lisa Hales for helpful discussions and clarifications on the
subject of these notes.

\end{document}